\documentstyle[12pt,aaspp4,astrobib,epsfig]{article}
\newcommand{\lapprox }{{\lower0.8ex\hbox{$\buildrel <\over\sim$}}}
\newcommand{\gapprox }{{\lower0.8ex\hbox{$\buildrel >\over\sim$}}}
 
\newcommand{\hi}{\mbox {H\,{\sc i}}}                  

\def\h2o{H$_2$O}
\def\kms{km s$^{-1}$}

\def\mone{$^{-1}$}
\def\mtwo{$^{-2}$}
\def\mthree{$^{-3}$}
\def\Msun{\hbox{${\rm M}_{\sun}$}}

\def\h2o{H$_2$O}
\def\d21{D_{21}}

\def\mc#1{\multicolumn{1}{c}{#1}}
\def\mcc#1{\multicolumn{2}{c}{#1}}
\def\mccc#1{\multicolumn{3}{c}{#1}}

\def\n11{n_{11}}
\def\n10{n_{10}}
\def\t400{T_{400}}
\def\f1{f_{(-2)}}

\def\e4{\epsilon_{(-4)}}

\lefthead{Gallimore et al.}
\righthead{\hi\ Absorption in Seyfert Galaxies}
\begin{document}

\title{Neutral Hydrogen (21~cm) Absorption in Seyfert Galaxies: Evidence for
  Free-free Absorption and Sub-kiloparsec Gaseous Disks}   
\author{J.F. Gallimore}
\affil{National Radio Astronomy Observatory\footnote{The National
    Radio Astronomy Observatory is operated by Associated
    Universities, Inc., under contract with the National Science
    Foundation.}\\ 520 Edgemont Rd.\\Charlottesville, VA
  22903\\(jgallimo@nrao.edu)} 

\author{S.A. Baum, C.P. O'Dea}
\affil{Space Telescope Science Institute, 3700 San Martin Dr.\\
Baltimore, Maryland 21218}

\author{A. Pedlar}
\affil{NRAL, Jodrell Bank\\Macclesfield, Cheshire SK11 9DL\\UK}

\and
\author{E. Brinks}
\affil{Departamento de Astronom\'{\i}a\\Universidad de
  Guanajuato, Apdo. Postal 144\\ Guanajuato, C.P. 36000, Mexico}

\begin{abstract}
Active galaxies are thought to be both fueled and obscured by neutral
gas removed from the host galaxy and funneled into a central accretion
disk.  We performed a VLA imaging survey of 21~cm absorption in
Seyfert and starburst nuclei to study the neutral gas in the
near-nuclear environment. With the exception of NGC~4151, the
absorbing gas traces 100~pc scale, rotating disks aligned
with the outer galaxy disk. These disks appear to be rich in atomic
gas relative to nuclear disks in non-active spirals.  We find no
strong evidence for rapid in-fall or out-flow of neutral hydrogen, but
our limits on the mass infall rates are compatible with that required
to feed a Seyfert nucleus.

Among the galaxies surveyed here, neutral hydrogen absorption traces
parsec-scale gas only in NGC~4151. Based on the kinematics of the
absorption line, the disk symmetry axis appears to align with the
radio jet axis rather than the outer galaxy axis (cf.  \citeNP{MHPMKA95}). The most surprising result is that we detect no 21~cm
absorption towards the central radio sources of the hidden Seyfert 1
nuclei Mrk~3, Mrk~348, and NGC~1068.  Moreover, 21~cm absorption is
commonly observed towards extended radio jet structure but appears to
avoid central, compact radio sources in Seyfert nuclei.  To explain these
results, we propose that 21~cm absorption towards the nucleus is
suppressed by either free-free absorption, excitation effects (i.e.,
enhanced spin temperature), or rapid motion in the obscuring
gas. Ironically, the implications of these effects is that the
obscuring disks must be small, typically not larger than a few tens of
parsecs.
\end{abstract}

\keywords{galaxies: active: galaxies: Seyfert, galaxies; radio lines:
  galaxies} 

\section{Introduction}\label{intro}
\nopagebreak
The commonly accepted model explaining the power source for active
galactic nuclei (AGNs) is a massive black hole fed by an accretion
disk. The advantage of this model is the high efficiency of conversion
of gravitational energy to radiant energy, but the main difficulty is
the removal of angular momentum. Unlike stars, which constitute a
dissipationless medium, gas and dust can lose orbital energy and
angular momentum through collisions and shocks.  At least in Seyfert
galaxies, but probably in all AGNs, the host galaxy reserves a mass of
neutral ISM comparable to or exceeding the lifetime needs of the
AGN. Except in galaxy clusters, the mass of the hot phase component of
the ISM is insufficient to fuel the most luminous AGNs over their
duration. Although the transport mechanism remains a mystery, the
conventional model holds that AGNs are powered by neutral ISM removed
from the host galaxy (e.g., \citeNP{Rees84}).

Important variants on the standard model are the unifying schemes for
AGNs. The dust associated with the neutral ISM can obscure a range of
sight-lines to the AGN, affecting its outward appearance. It is
proposed that both broad-line (e.g., Seyfert 1) and narrow-line
(Seyfert 2) AGNs are identical, but the broad-line region (BLR) of
Seyfert 2 galaxies may be obscured by a dusty ring, viewed more nearly
edge-on, surrounding the active nucleus (this model is reviewed in,
e.g., \citeNP{Lawrence87}; \citeNP{Antonucci93}; \citeNP{Ward96}).  There are a handful of hidden Seyfert~1 galaxies discovered
by spectropolarimetry (e.g., \citeNP{AM85};
\citeNP{MG90}; \citeNP{TMK92}), but
it is not clear whether the model is generally applicable. The
detection of relatively high X-ray obscuring columns towards Seyfert 2
nuclei warrants the question of the nature and scale of the obscuring
medium (e.g., \citeNP{AKIH91}; \citeNP{MMW92}; \citeNP{MCWMW93}; \citeNP{SD96}; \citeNP{TGNM97}).

We have observed a sample of 13 radio-bright Seyfert galaxies in \hi\
(21~cm) absorption to explore the distribution and kinematics of the
neutral ISM in the vicinity of the AGN. There are two primary
advantages of this technique. First, absorption lines are potentially
more sensitive than emission lines, the former limited primarily by
the brightness of the background source rather than the column density
or emission measure of the gas. Second, one can use powerful aperture
synthesis arrays to resolve the absorption line gas on scales as small
as the background radio source (i.e., potentially down to sub-parsec
scales). Conversely, although the absorption experiment may not
resolve the background source, the structure of the background source
may be known from independent continuum studies, and so the effective
resolution of the absorption experiment is improved. A third advantage
is that one knows for certain that the absorption line gas must lie in
front of the background source, removing the ambiguity suffered by
emission line studies. Of course, the primary limitation is that one
can detect and resolve absorption line gas only to the extent of
the background source. In addition, because the opacity is inversely
proportional to the excitation temperature, absorption lines are
weighted more heavily by colder clouds. 

We present this work as follows. Section~\ref{obs} describes the
sample selection, observations, and data reduction
techniques. Section~\ref{results} presents the data and detection
statistics. In Section~\ref{props} we discuss the properties of the
\hi\ absorbing gas, including quantitative treatments of the spin
temperatures and \hi\ column densities. We also compare the observed
\hi\ columns with hydrogen columns based on X-ray
spectroscopy. Section~\ref{models} describes the distribution and
kinematics of the \hi\ absorption based on the application of rotating
disk models. Sections~\ref{schemes} and \ref{fueling} discuss the implications of
these observations on the standard model for AGNs, and we summarize
our main conclusions in Section~\ref{conclusions}.

\section{Observations and Data Reduction}\label{obs}
\nopagebreak

\subsection{Sample Selection}

The selection criteria for this survey were spiral galaxies with
evidence for nuclear activity and radio brightness $S_{\nu} > 50$~mJy
as resolved by the VLA A-Array. In essence this survey is a radio
flux-limited sample comprising mainly Seyfert galaxy nuclei. Our
original survey consisted of 24 active spirals, but owing to a
technical problem with our first observation (VLA program AB~605 on 19
Jul 1991), we were unable to recover the \hi\ spectra for these
sources. We received a reduced time allocation for a repeat
observation on 22 Nov 1992. Time and LST constraints reduced our
sample to a subset of 13 of the brightest sources (VLA program
AB~658).

\subsection{Observations}

We observed the sample galaxies with the VLA operating in its
A-configuration, i.e., the longest baseline configuration, with
baselines ranging from $\sim 3.2$--170~k$\lambda$ at 21~cm. The
spatial resolutions in the image plane are typically $\sim
1\farcs5$--2\arcsec.  A summary of the main observing parameters are
provided in Table~\ref{obsparms}.  Receivers were tuned to the
appropriately redshifted wavelength of the \hi\ 21~cm line (rest
frequency $\nu = 1420.406$~MHz). We used the optical definition for
velocity and referenced the velocities to the heliocentric
frame. Eleven sources were observed in 4-IF mode, with paired IFs
tuned redward and blueward of the systemic velocity to expand the
velocity range covered by the receivers. The bandwidths are 3.1~MHz
divided into 32 channels. In velocity units, the channel widths are
$\sim 21$~\kms, and, accounting for channel overlaps between the IF
pairs, the total bandwidth is $\sim 1050$~\kms.  The two remaining
sources (NGC 2992 and NGC 5506) were observed using 1A mode covering
a single bandpass of 6.25~MHz, again with channel widths of $\sim
21$~\kms. 

Source observations interleaved with shorter scans of a point-source
calibrator for phase referencing. Each source observation was
accompanied by a scan of a bright flux standard, either 3C~48 or
3C~286. We also used these sources for bandpass calibration to remove
the residual channel-to-channel complex gain variations.

\subsection{Data Reduction}

We reduced the data using standard procedures in AIPS. The flux scale
was set by scans of the radio continuum standards 3C~48 and 3C~286,
defined according to \citeN{BGPW77}. Absolute calibration of
interferometric phase was determined by point source models for the
phase reference calibrators. We then employed 2--3 iterations of
self-calibration on each of our targets, properly including all of the
confusing sources brighter than 5~mJy within two primary beam
widths. The final CLEAN-deconvolved models were subtracted from the
$(u,v)$ database. We generated spectral-line cubes for each target by
Fourier transform. The cubes were inspected for residual continuum
emission or bandpass gradient not subtracted by the CLEAN-component
model. For stronger sources, we subtracted a linear fit to the
spectral baseline, but for weaker sources it sufficed to remove a
continuum mean. All residual subtractions were performed in the image
(i.e., not $u,v$) plane.

A possible source of error in this experiment is a bandpass
calibration artifact resulting from temperature variations, and hence
length variations, of the waveguide system at the VLA (\citeNP{Carilli91}; \citeNP{Lilie94}). The signature is a sinusoidal ripple with
a period of 3~MHz, matching the bandwidth used in this
experiment. Among the sources observed for this experiment, only the
nuclear spectrum of NGC~1068 displayed broad spectral features that
could be explained by the 3~MHz ripple. To test for this artifact, we
generated a spectrum of the phase calibrator for NGC~1068. The phase
calibrator showed similar spectral structure, corresponding to a
spectral dynamic range of $\sim 400$. We therefore have little
confidence in the broad spectral features of the nuclear 21~cm
absorption profile of NGC~1068.

\section{Results}\label{results}

We detect \hi\ absorption towards 9 out of the 13 sources observed in
this survey, making a detection rate of 69\%. The breakdown by nuclear
activity is 3 out of 4 Seyfert 1 nuclei detected (Fig.~\ref{sy1spec});
4 out of 6 Seyfert 2 nuclei detected (Fig.~\ref{sy2spec}); and 2 out
of 3 starburst/LINER nuclei detected (Fig.~\ref{sbspec}). Explicitly,
the non-detections are Mkn~668 (Seyfert 1), Mkn~3 and Mkn~348 (both
Seyfert 2 with hidden Seyfert 1 nuclei), and NGC~2639 (LINER). We
summarize the basic properties of the integrated \hi\ absorption in
Table~\ref{hiprops}. Parameters of the integrated absorption line
profiles are provided in Table~\ref{hikin}.

In several sources, the radio continuum was sufficiently resolved that
we could isolate or map the \hi\ absorption distribution against the
extended continuum. Sources in which the \hi\ absorption is
unresolved, but the radio continuum is resolved, are NGC~2110
(Fig.~\ref{n2110tot}), NGC~2992 (Fig.~\ref{n2992tot}), NGC~4151
(Fig.~\ref{n4151tot}), and Mkn~6
(Fig.~\ref{mk6tot}). The continuum and \hi\ absorption in NGC~1068 are
resolved (Figs.~\ref{n1068tot} and \ref{n1068panel}), and the
\hi\ absorption in NGC~3079 is marginally resolved against the
background continuum (Figs.~\ref{n3079tot} and \ref{n3079panel}).
Follow-up MERLIN observations have better resolved the \hi\ absorption
towards NGC~3079 \cite{PMGBO96}, NGC~4151 \cite{MHPMKA95}, and Mkn~6 \cite{GHPM98}. \hi\ absorption
towards Mkn~231 has been resolved by the VLBA \cite{CWU97}. Appendix~\ref{individuals} details more general
characteristics of these sources.

The only published survey comparable to the present one is Dickey's
(1986) \nocite{Dickey86} non-imaging VLA survey of radio bright
spiral and irregular galaxies. Accepting the small number statistics,
our detection rate (69\%) is better than Dickey's (53\%). The most likely
explanation is a $\sim 4\times$ improvement in sensitivity rather than
sample selection. For example, our survey contains a greater fraction
of AGNs relative to star-forming nuclei, but isolating the subsample
of AGNs within Dickey's survey does not increase the detection rate.
Explicity, nineteen sources were surveyed in \citeN{Dickey86},
including 11 Seyfert or LINER nuclei and one BL Lac; rapid
star-formation may dominate the radio continuum emission in the
remaining sources.  Dickey detected 10 out of 19 (53\%) galaxies, or
$< 78\%$ accounting for 5 additional, but marginal, detections. The
statistics are comparable for the subsample of Seyfert and LINER
nuclei: 5 ($+4$ marginal) out of 12 were detected, for a detection
rate of 42\% ($<75\%$). Furthermore, the distribution of continuum
source brightnesses are similar between the present survey and
\citeN{Dickey86}. The improved flux sensitivity of the present
survey constitutes an improvement in optical depth sensitivity.  An
example is the starburst nucleus NGC~3504, which was detected
marginally in \citeN{Dickey86} but is clearly detected in our
survey.

\section{Properties of the \hi\ Absorption Line Gas} \label{props}

\subsection{Spin Temperatures}\label{tspin}

The column density $N_{HI}$ and spin (excitation) temperature $T_S$ are
related to the observed \hi\ absorption spectrum according to:
\begin{equation}
N_{HI} = 0.182\times 10^{21}\ {\rm cm^{-2}}\ (T_S/100~{\rm\ K})\ \int
\tau_{v}\ dv\ ,
\label{eq-nh}
\end{equation}
where $v$ is in units \kms, and $\tau_v$ is the opacity measured against
the background continuum. In order to derive \hi\ column densities, we
need some handle on the excitation state of the absorbing gas.

There are three factors controlling the spin temperature of neutral
hydrogen (\citeNP{Field59a}; \citeNP{Field59b}; \citeNP{Field59c}):
collisions, pumping by Ly-$\alpha$ photons, and the absorption of
21~cm continuum radiation. The spin temperature approaches the kinetic
temperature $T_K$ in collisionally dominated (high density) regions.
\hi\ absorption columns through dense clouds in galaxy disks typically
match the \hi\ emission columns for a spin temperature of $T_S \sim
50$--300~K, close to the expected $T_K$ (e.g., \citeNP{DTS78}; \citeNP{PST83};
\citeNP{LvdHBO83}; \citeNP{Braun97}). However, radiative effects
are expected to be important in the neighborhood of an AGN,
\cite{BE69}, and $T_K$ is a {\em lower} limit for
$T_S$.

Fig.~\ref{fig-tspin} demonstrates the effects of 21~cm absorption and
Ly-$\alpha$ pumping of the ground state of \hi\ in the 100~pc scale
environment of a Seyfert nucleus (compare with Fig.~5 in
\citeNP{OBG94}). For illustration, we modeled
the 21~cm radiative effects using the nuclear properties of NGC~4151.
The background radio sources have a brightness temperature $T_R \sim
10^6$~K at 21~cm (e.g., \citeNP{HPUBGP86};
\citeNP{URCW98}), and the specific luminosity of the
continuum near Ly-$\alpha$ is $F_{Ly\alpha} \approx 1.8 \times
10^{27}$ ergs s\mone\ Hz\mone\ \cite{Ketal92}.  In computing
the density of Ly-$\alpha$ photons we further assumed a distance of
50~pc between the continuum source and the absorbing \hi\ cloud. A
distance of 50~pc usually falls within the range of the observed,
deprojected distances between the radio nuclei and \hi\ absorbing
regions (see \S\ref{models}). We find that, in the 100~pc vicinity of
an AGN, Ly-$\alpha$ pumping or 21~cm absorption controls $T_S$ for
densities $N_{HI} \la 1000$~cm\mthree, but $T_S$ settles
asymptotically to $T_K$ for $n_{HI} \ga 1000$~cm\mthree\ as collisions
dominate the excitation of the ground state.

Note that Fig.~\ref{fig-tspin} is representative for typical
Seyfert jets. A few Seyferts, however, have a bright self-absorbed
core; Mrk~348 is an example. The intense radiation field near a
self-absorbed radio core strongly affects $T_{spin}$ (e.g., Bahcall \&
Ekers 1969), and this effect may suppress \hi\ absorption from an
obscuring torus. We consider this possibility in more detail in
\S\ref{exciteme}. 

For most of the Seyferts in this survey, the \hi\ absorbed regions
cover radii comparable, in projection, to the size of the NLR, and \hi\  
absorption might plausibly arise from neutral filaments embedded
within or shadowed by NLR plasma.  Since $\tau_v \propto (T_S\ \Delta
v)^{-1}$, \hi\ absorption studies are particularly sensitive to gas that is
cold both thermally and kinematically. We would therefore expect most
of the \hi\ absorption to arise from the coldest regions of clouds
where hydrogen is primarily atomic. In AGN-irradiated clouds, the
kinetic temperature of the cold atomic region is $T_K \approx
100$~K \cite{MHT96}. If we assume pressure
equilibrium and typical NLR conditions $T_e \sim 15000$~K and $n_e
\sim 2000$~cm\mthree\ \cite{Koski78}, the density of the cold atomic
region is $n_{HI} \sim 10^5$~cm\mthree. At this limit collisions
dominate, and $T_S \rightarrow T_K$ (Fig.~\ref{fig-tspin}).

We conclude that \hi\ absorption from the NLR region of Seyfert
galaxies probably arise from the cold dense regions where $T_S \approx
T_K \sim 100$~K. This result is remarkable in that $T_S$ in cold
Galactic clouds is also $\sim 100$~K, and so the conventional
assumption $T_S = 100$~K holds up under a broad range of conditions.
Of course, there may be situations where \hi\ absorption traces a
warmer atomic medium in the AGN environment; for example, the
sight-line to the radio source may not cross the cold neutral region.
In the warm atomic region $T_K$ (and $T_S$) can elevate to several
1000~K before hydrogen is significantly ionized \cite{MHT96}. The assumption of $T_S = 100$~K
therefore provides a lower limit to the true \hi\ column.

\subsection{Column Densities}\label{nh}

We derived the column densities of the absorbing \hi\ according to
Eqn.~\ref{eq-nh} and assuming a spin temperature $T_S = 100$~K.  The
measured columns are listed in Table~\ref{hiprops}.  The absorption
opacity $\tau$ was measured against channel-averaged continuum images
from the same data set.  Confusion between continuum sources can
dilute the measured opacities or, equivalently, the \hi\ absorbing
gas might not completely cover an unresolved background source. The
reported values of $N_{HI}$ are therefore lower limits to the true
absorbing column.  The columns for Mkn~6, NGC~4151, and Mkn~231 are
derived from published MERLIN and VLBA observations (\citeNP{MHPMKA95}; \citeNP{CWU97};
\citeNP{GHPM98}). 

One concern of this study is whether the \hi\ absorbing gas is
peculiar to the small-scale AGN environment or arises from the disk of
the host galaxy on larger scales. Addressing this question, a simple
prediction would be that the \hi\ absorption should show no trend with
inclination, since AGN disks appear to be oriented randomly with
respect to outer galaxy disks (e.g., \citeNP{SKSA97};
\citeNP{NWMG99}).  Fig.~\ref{nh-vs-inc} compares the
distribution of \hi\ columns with the inclination of the host galaxy,
estimated from optical axial ratios tabulated by \citeN{rc3} (compare with Fig.~5 in \citeNP{Dickey86}).
Fig.~\ref{nh-vs-inc} also includes \hi\ absorption measurements from
this work and \citeN{Dickey86}. Only those sources with
well-defined disk morphology and optical axial ratio are plotted.  For
reference, we also plot the range of Galactic \hi\ absorbing columns
observed towards extragalactic radio sources \cite{PST82} and the radio source Sgr~A in the Galactic Center
\cite{SEG82}. 

We can draw two conclusions from Fig.~\ref{nh-vs-inc}.  First, running
counter to the prediction for AGN disks, there is evidence for a trend
in \hi\ column with host galaxy inclination.  For instance, all four
sources with $i > 60\arcdeg$ are detected in \hi\ absorption.  We
computed correlation statistics for the regression of $N_{HI}$ on
$\sec{i}$ using standard survival analysis techniques to account for
the non-detections. The results indicate a real trend in $n_{HI}$ with
$i$: the probability for no correlation is 2.7\% for Kendall's
generalized $\tau$, $< 1\%$ for Cox's proportional hazard model, and
4.6\% for Spearman's generalized $\rho$ (although we caution that
Spearman's $\rho$ is not reliable for sample sizes $\la 30$). This
result suggests that in many Seyferts \hi\ absorption arises in
gaseous disks at least aligned with, if not cospatial with, the
kiloparsec-scale galaxy disks. We caution that the trend is only valid
for the quantity $N_{HI} / T_{spin}$, but a significant spread in
$T_{spin}$ would reduce confidence in the trend of $N_{HI}$ with
inclination. Nevertheless, we demonstrate in \S\ref{models} that the
kinematics and distribution of the \hi\ absorption line gas are also
consistent with the observed trend.

The second conclusion is that the typical absorbing columns towards
detected Seyfert galaxies exceed the Galactic columns by factors ranging
from $\sim 3$--30, although there is some overlap for individual
Seyfert galaxies and Galactic absorbing clouds. The mean face-on
column for detected Seyferts is $4\times 10^{21}$~cm\mtwo, exceeding
Galaxy columns by a factor of nearly 10. It would be interesting to
compare the Seyfert \hi\ columns with a larger sample of non-active
galaxies, but, lacking bright radio sources, non-active galaxies are
not readily studied in \hi\ absorption. Nevertheless, based on \hi\ 
emission studies, the face-on columns through spiral galaxies are
typically a few times $10^{20}$~cm\mtwo\ (e.g., surveys include the
theses of \citeNP{Bosma78}; \citeNP{Begeman87};
\citeNP{drielthesis}; their associated publications; and
\citeNP{BvW94}).  Typical \hi\ emission columns
of normal spiral galaxies are comparable to the absorption columns
measured for the Galaxy rotated to a face-on view.

The observed enhancement of \hi\ column towards Seyfert nuclei is
probably not an excitation effect. Demonstrated above in
\S\ref{tspin}, the measured columns are lower limits because 21~cm
absorption and Ly-$\alpha$ pumping in the vicinity of an AGN can raise
$T_S$ to 1000~K or more. In contrast all of the Galactic values have
been corrected for directly measured spin temperatures (for the
Galactic Center column we assumed $T_S \approx T_K \sim 300$~K in the
10~pc diameter disk after \citeNP{GWCT85} and references
therein). Furthermore, owing to proximity, \hi\ absorbing clouds in
the Galaxy are better resolved compared to \hi\ absorption towards
Seyfert nuclei.  Therefore, the Galaxy columns suffer less continuum
dilution owing to confusion or covering fraction arguments. We
conservatively conclude that the \hi\ absorption columns towards
Seyfert nuclei exceed the \hi\ columns in the Galaxy, and perhaps
other non-active spiral galaxies, by a factor of at least six
(as shown in Fig.~\ref{nh-vs-inc}). 

We show in \S\ref{models} that \hi\ absorption in Seyfert galaxies
usually traces \hi\ disks in the inner kiloparsec. In contrast, \hi\ 
columns are often reduced below detectability in the inner
kiloparsec of nearby, non-active spirals (e.g., \citeNP{Roberts75p309};
\citeNP{GH88p522}). Limited by instrumental
sensitivity, the central columns are typically $\la C_f\ 
10^{20}$~cm\mtwo, where $C_f$ is the beam covering fraction of the
\hi\ gas. The high detection rate for Seyfert nuclei argues for $C_f$
approaching unity. Based on the interpretation of \S\ref{models}, it
appears that the \hi\ surface density in the inner kiloparsec of
Seyfert and starburst galaxies exceeds by a factor $\ga 20$ the \hi\ 
columns in the centers of non-active galaxies. We discuss this result
further in \S\ref{fueling}. 

\subsection{Comparison with X-ray Columns}\label{xray}

X-rays provide another measure of the column of material obscuring the
central engine in active galaxies. Photoelectric absorption by
foreground gas that is not fully ionized attenuates the soft X-ray
continuum spectrum of the central engine. The simplest models for AGN
X-ray spectra usually involve a power law continuum modified by
photoelectric absorption through the Galaxy and photoelectric
absorption intrinsic to the source. The absorption is commonly
parameterized by a total hydrogen column adopting standard cross sections
and abundances (e.g., \citeNP{MM83}). 

Table~\ref{tab-cx} lists the X-ray absorption columns available in the
literature for sources in the combined survey. There are often several
models for a given source, and predictably the absorption column
becomes poorly constrained as models become more complex. In each
case, we selected the simplest model that provides a reasonable fit
and the best constraints on an absorption column. Selecting models in
this way tended to err on the side of caution; that is, more
sophisticated models often give larger measures for the X-ray
photoelectric column. The conclusions of this section are not affected
by the selection of the simplest X-ray spectral models.

Fig.~\ref{fig-c21cx} compares the 21~cm absorption columns with the
X-ray absorption columns. Under the hypothesis that the radio and
X-ray nuclei are effectively cospatial, the prediction is a
correlation between the 21~cm and X-ray columns. In addition, one
would also expect that the X-ray columns should tend to be slightly
greater than the 21~cm columns, since 21~cm absorption is not
sensitive to the presence of hydrogen locked in molecules or partially
ionized gas.

We find no measurable correlation between 21~cm and X-ray absorption
columns, and the columns match reasonably only for Mrk~231. For most
(8 out of 11) of the sources, the X-ray column exceeds the 21~cm
column by at least an order of magnitude. We refer to these sources as
having a ``photoelectric excess'' of absorption. There are three
explanations for photoelectric excesses. First, it may be that most of
the hydrogen is in molecular or ionized form. Alternatively, it may be
that the primary X-ray absorber is dust, and the obscuring material
has a lower gas-to-dust ratio than assumed in the modeling. The former
explanation seems most likely for the Seyfert 1 nuclei, in which it
appears that the absorbing medium is ionized (e.g., \citeNP{GTNNMY98}; \citeNP{FBEFIM99}). The second explanation is
that the radio emission may not be cospatial with the nuclear X-ray
source, and so X-ray absorption and 21~cm absorption may probe
different regions of the foreground media. We consider this
possibility further in
\S~\ref{schemes}. The third explanation might be spin temperature
effects. Increasing $T_{spin}$ to $\sim 1000$~K would produce a better
agreement between X-ray and 21~cm columns for Mrk~6, NGC~2110,
NGC~4151, and NGC~5506, and the remaining photoelectric excess sources
require $T_{spin} > 10^4$~K to explain the discrepancy. This
explanation fails, however, to explain the result that \hi\ absorption
tends to avoid the nucleus in those Seyferts where the radio jet is
resolved (\S\ref{models}, below). It seems more likely that 21~cm
absorption does not trace the same gas giving rise to the soft X-ray
opacity. 

The sources NGC~2992 and NGC~3079 show a ``photoelectric deficit,''
meaning that the 21~cm absorption column exceeds the X-ray
photoelectric column. Again, it may be that the abundances assumed for
the X-ray modeling are incorrect. For example, in this case, the
gas-to-dust ratio may be {\em higher} than was assumed for the X-ray
spectral modeling. The host galaxies of both sources are edge-on, and
both show evidence for extended soft X-ray emission (e.g.,
\citeNP{EFWB90}; \citeNP{RPS97};
\citeNP{CBOV98}). The spatial resolution of the soft
X-ray spectra is very coarse, $\sim30\arcmin$, much larger than the
VLA beam.  An alternative explanation, then, might be that extended
X-ray emission dilutes the soft X-ray columns towards the nucleus. The
extended emission that is resolved by current X-ray satellites is only
$\sim 10\%$--20\% of the total, insufficient to explain the
photoelectric deficits. Dilution of the soft X-ray columns would
require extended X-ray emission on sub-arcminute scales not yet
resolved; future X-ray satellites will have sufficient resolution to
test the hypothesis.

We conclude that there is no correlation between X-ray photoelectric
columns and 21~cm columns. Although there are a number of possible
explanations, the simplest explanation may be that the radio and X-ray
sources are not cospatial, and so 21~cm absorption and soft X-ray
absorption do not (usually) trace the same foreground gas. We consider
the implications in more detail in the following discussion
(particularly \S\ref{schemes}, below).

\section{Disk Models for the Distribution and Kinematics of the \hi\
  Absorption}\label{models}  

There are now a handful of Seyfert nuclei in which the \hi\ absorption
is resolved against the background radio source: Mkn~6 (this study;
\citeNP{GHPM98}), Mkn~231 (\citeNP{CWU97}), NGC~1068 (this study, Fig.~\ref{n1068tot};
\citeNP{GBOBP94}), NGC~2110 (this study,
Fig.~\ref{n2110tot}), NGC~2992 (this study, Fig.~\ref{n2110tot}),
NGC~4151 (this study, Fig.~\ref{n4151tot}); \citeNP{MHPMKA95}), \& NGC~5929 (\citeNP{CPMGH98}). \hi\ absorption is
also resolved in the starburst nucleus of NGC~3079 (this study;
\citeNP{PMGBO96}; \citeNP{SIN98p219}). The
remaining detections that we have imaged, but in which the \hi\ 
absorption is not well resolved against the background continuum, are
NGC~3504 and NGC~5506.

It is remarkable that the \hi\ absorption tends to avoid the radio
nucleus. More typically, the absorption is displaced by $\ga 100$~pc
in projection away from the central radio source
(Table~\ref{hiprops}). The exceptions are NGC~3079 and NGC~5506, which
are edge-on, and NGC~4151, in which the \hi\ absorbing gas lies within
a few tens of parsecs of the central engine \cite{MHPMKA95}. We have demonstrated that detections are more likely
with increasing inclination of the host galaxy (\S\ref{nh};
Fig.~\ref{nh-vs-inc}; see also \citeNP{Dickey86}). Moreover, the
inferred absorption columns are not correlated with estimates of the
X-ray absorption column (\S\ref{xray}). It appears that \hi\ 
absorption in Seyfert and starburst nuclei traces gas in the inner
disk of the host galaxy rather than gas associated with the AGN. In
particular, the \hi\ absorption does not appear to be associated with
a parsec-scale obscuring torus, with NGC~4151 being an important
exception.

To test this conclusion, we modeled the kinematics of the \hi\ 
absorbing gas as disks in circular rotation. We spatially
resolve the kinematics only in NGC~1068 \cite{GBOBP94},
and so had to fit disk models to global absorption line profiles rather
than spatially resolved channel maps. The results are listed in
Table~\ref{t-diskfits} and displayed in
Figs.~\ref{mkn6norm}--\ref{n5506norm}. 

\subsection{Procedure}

The disk models were computed and fit as follows. First, we generated
a background continuum model from the continuum CLEAN components. For
NGC~4151 we constructed a continuum model based on VLBI observations
\cite{URCW98}, which recover all of the continuum flux
in the MERLIN maps around the region of the \hi\ absorption
\cite{MHPMKA95}. We next generated a model disk described
by vertical column density $N_H$; rotational velocity $v_{rot}$;
velocity dispersion $\sigma_v$, fixed at $\sigma_v = 10$~\kms, as is
commonly observed in spiral galaxy disks (e.g., \citeNP{SvdK84}); scale height $h$, fixed at 10~pc, self-consistent
with the measured rotation velocities and disk radii; and,
particularly for edge-on galaxies, inner and outer annular radii
$r_{in}$ and $r_{out}$. The scale height $h$ for NGC~4151 was fixed at
0.1~pc because of the small radial scale of the absorbing disk. We
also fitted the rotation curve as a power law $v \propto r^{\alpha}$,
but, in every case, $\alpha \la 0.01$, meaning an essentially flat
rotation curve over the measured region.

Since the hypothesis to be tested is whether the \hi\ absorption
traces normal disk gas, the inclination $i$ and kinematic major axis
$\phi$ were fixed to those values observed in the outer disk of the
host galaxy in question. For NGC~1068, we took into account the
rotation of the kinematic axis by 45\arcdeg\ between the outer and
inner galaxy. The apparent cause of this rotation is streaming in the
central stellar bar \cite{GBOBP94}.  Lacking spatial
information, except for NGC~1068, we further assumed that the
absorbing \hi\ uniformly fills the disk. In each case where the \hi\
absorption is displaced from the nucleus, we assumed that the near
side of the disk lies on the same side of the nucleus as the offset
absorption. The (deprojected) extent of the observed
\hi\ absorption determines the inner and outer radii of the disk. For
the edge-on galaxies in our survey, only the line profile constrains
the inner and outer radii of the absorbing disk.

Having generated and properly oriented the model disk and background
radio source, we then integrated the line-of-sight opacity $\tau(v)$
and computed an absorption spectrum for each background CLEAN
continuum component. The final spectrum was created by adding up all
of the model absorption spectra. We fit the model spectrum against the
observed \hi\ absorption spectrum using a downhill simplex algorithm
to minimize $\chi^2$ (chapter 10.4 of \citeNP{PTVF92} and
references therein). The best-fit model was then checked for
self-consistency against the observed \hi\ absorption morphology. We
performed this check to ensure that the model disk would not obscure
unabsorbed radio structure.

After convergence, we estimated standard errors using a Monte Carlo
technique. We assumed that the best-fit model line profile was the
correct profile, and added random, Gaussian noise to the model
spectrum to simulate an observed spectrum. The r.m.s. of the simulated
noise was set equal to that of the residual spectrum (i.e., data $-$
model). We repeated the fitting procedure on a series of ten simulated
line profiles. The standard errors were derived from the deviations of
parameters fitted to the simulated spectra. The failing of this
technique is that it does not properly account for real absorption not
present in the model and so may underestimate the real uncertainties
where there are significant residuals. On the other hand, the measured
standard errors appear to reflect the quality of the models and may
not be far from their true values. 

The models are not unique in the sense that one can always contrive a
model where all of the gas is falling into or flowing away from the
background radio source. On the other hand, the structure of the
background radio source leaves a ``fingerprint'' on the absorption
profile that promotes the uniqueness of the fitted parameters of the
disk model.  That is, the structure of the background source will
preferentially enhance or suppress the contribution of certain
velocities to the spatially integrated line profile. An important
failing is that our models cannot fit for spatially varying opacity in
the absorbing gas because plausible opacity variations, such as spiral
arm structure, require too many parameters for the fits to be
meaningful. Nevertheless, the uniform disk model produces a reasonable
and illustrative fit for all of the observed line profiles.

One potential artifact of the fitting is a parsec-scale inner radius
for the absorbing disk. This artifact results mainly from the need to
fit the high velocity wings of the absorption line profile, since, for
a given annular section of the disk at radius $r$ surrounding a radio
source of size $r_{0}$, the velocity width is $\Delta v \sim v_{rot}
[r_{0} / \sqrt{(r-r_0)^2 + r_0^2}]$.  The need for a small inner
radius might be relaxed by allowing larger velocity dispersions or
streaming velocities. We chose to keep the dispersions fixed, since
often the dispersions would grow unrealistically large in order to fit
a discrete velocity component (e.g., NGC~1068, Fig.~\ref{n1068norm};
and NGC~5506, Fig.~\ref{n5506norm}). Fitted values for the disk inner
radius should be viewed as a lower limit.

\subsection{Results}

With the exception of NGC~4151, which is described further below, the
disk models generally describe the observed \hi\ absorption profiles and
locations very well. Poor values of $\chi^2$ can be traced generally
to a discrete velocity component for which the simple disk model
cannot account. Any discrete, residual components always arise within
the velocity range traced by the model disk profile. We therefore
suspect that unresolved opacity variations in the foreground disk may
give rise to the discrete features in the line profile. For example,
the spectrum of NGC~5506 shows a discrete redshifted component
(Fig.~\ref{n5506norm}) that might arise from a cold cloud in the main
disk which happens to fall along the line-of-sight.

To demonstrate the validity of the galaxy disk model, we point to the
peculiar absorption line profile of NGC~2110 (Fig.~\ref{n2110norm}).
The absorption line is redshifted from systemic by nearly 300~\kms,
and the traditional interpretation would be infall towards the galaxy
nucleus. However, the VLA images reveal that the absorbing gas is
displaced from the nucleus along the kinematic major axis. Because of
the special, chance alignment of the kinematic major axis and the
radio jet, the northern lobe of the jet back-illuminates receding disk
gas, and the result is a redshifted line profile. The uniform disk
model reasonably reproduces the line profile shape, lending
credence to this interpretation. The lack of a blue-shifted absorption
line against the opposing, southern radio lobe indicates that the jet
points in front of the disk to the south.

NGC~4151 represents a special case, since the \hi\ absorption probably
traces gas within the inner few tens of parsecs of the nucleus.
Unfortunately, the location of the nucleus in the radio reference
frame is not known, since all of the compact continuum sources are
steep spectrum (e.g., comparing \citeNP{PKLMABOU93} and
\citeNP{CTH90}). Since the \hi\ absorbed component
(C4) is also free-free absorbed, and since the radio jet and NLR
probably point toward us west of C4, the nucleus is probably near and
west of C4 \cite{PFHRD98}. We selected the central VLBI
component C4W as the location of the nucleus after \citeN{MHPMKA95}. Discussed below, this placement only affects the
quantitative results of the model but does not alter the qualitative
conclusions.

The radio jet of NGC~4151 does not align with the minor axis of the
host galaxy, and so the AGN disk and galaxy disks are also misaligned.
We therefore tried two disk models for NGC~4151, one with an
orientation matching the host galaxy \cite{PKLMABOU93}, and
one matching the inferred properties of an AGN disk \cite{MHPMKA95}. The AGN disk model fits the absorption spectrum better
than the galaxy disk model. The reason is that the \hi\ 
absorption line centers very near the systemic velocity, and the AGN
disk model places the \hi\ absorbing gas along the disk minor axis.
The galaxy disk model, on the other hand, predicts a slightly
redshifted absorption component, since the absorption occurs towards
the receding axis. As a result, the galaxy disk model requires
counter-rotation to fit the observed \hi\ profiles. These conclusions
hold so long as the radio jet originates at the kinematical center of
the galaxy; the kinematical center may be placed anywhere along the
radio jet and an AGN disk will always fit the absorption spectrum
better than the galaxy disk model.

\subsection{Radial Streaming}\label{streaming}

To look for evidence of infall that might fuel the AGN, we repeated
the disk model fits allowing for a radial component to the gas
velocity. The best-fit streaming velocities are listed in
Table~\ref{t-stream}. The trivial, but unrealistic, solution would
have been to set the rotational velocity to zero and assume purely
radial motions. To avoid the trivial solution, we initialized the
iterative fitting procedure with the parameters from the purely
rotational model and allowed the circular and radial velocities to
relax into a local minimum for $\chi^2$.

Although the error-bars are usually large relative to the fitted
values, all of the streaming velocities are radially inward towards
the nucleus except for NGC~1068 and NGC~4151.  The deprojected infall
velocities are typically a few tens of \kms.

Looking at the exceptions, the \hi\ absorbed region of NGC~1068 traces
gas mainly on the trailing side of the central stellar bar
\cite{GBOBP94}, over which the mean radial velocity of
gas is expected to be away from the nucleus (e.g., \citeNP{RHvA79}; \citeNP{ST80}).
Although the streaming velocities in NGC~1068 are very uncertain, the
implied outward streaming is consistent with the predicted bar
streaming. It is all but impossible to interpret the outflowing gas
motions in NGC~4151 since the absorption is not yet resolved against
the VLBI structure.

An important caveat is the main failing of these disk models; namely,
we cannot recreate the distribution of absorbing gas as a function of
galactic coordinates. A localized enhancement of foreground column
covering a finite range of radial velocities can mimic enhanced
velocity dispersion or streaming velocities. Three-dimensional
($\alpha$, $\delta$, and $z$) models of the spectral line cubes would
add enough information to allow at least a crude treatment on the
spatial distribution of \hi\ absorbing gas, but the present data
generally lack the spatial resolution for three-dimensional fits to be
effective. The fitted streaming velocities can be viewed as upper
limits to the true radial motions on $\sim 100$~pc scales.

Accepting this caveat, the upper limits on the mass deposition rates
constrain the luminosity that can be powered by the \hi\ disks.  One
concern, for example, is whether the upper limits on the infall rates
might be insufficient to power the central engine.  We have included
in Table~\ref{t-stream} the mass infall rates estimated by $\dot{\rm
  M}\simeq M_{HI}\ v_{rad} / r_{out}$; typical values are a few
$\times 0.1$~\Msun\ yr\mone. Among the \hi\ detections presented here,
the mean, infall rate is $0.18$~\Msun\ yr\mone.  This rate can
support a nuclear luminosity of $L \simeq 10^{45}\ (\epsilon /
0.1)$~ergs s\mone, where $\epsilon$ is the efficiency factor for
mass-energy conversion in the accretion disk. This luminosity defines
the traditional Seyfert/QSO luminosity boundary, and so the inferred
mass deposition rates are comparable to that required to fuel Seyfert
nuclei. We caution that this result does not constitute proof of
infalling gas on 100~pc scales. On the other hand, we may conclude
that the kinematics of the \hi\ absorption are compatible with the
standard model for AGNs, namely, that the active nucleus is fueled by
gas removed from the host galaxy disk.

We should point out that the nature of the \hi\ absorption in active
spirals appears to be very different from the \hi\ absorption in
active ellipticals. Apart from a substantially lower detection rate,
\citeN{vGKEELP89} found that the \hi\ absorption lines
were never blue-shifted, indicating either tangential motion or radial
infall towards the central radio source. In contrast, we find that the
kinematics of \hi\ absorption in Seyfert galaxies appear to be
dominated by rotation, with a weak infall component superimposed.

\subsection{Stability Consideration}

Rotating gaseous disks are unstable to internal collapse and star
formation where the disk exceeds a critical surface density
$\Sigma_c$. Neglecting, for now, the contribution of molecular gas and
ionized gas to the local surface density, the ratio of the neutral
hydrogen surface density $\Sigma_{HI}$ to the critical density is
\begin{equation}
\frac{\Sigma_{HI}}{\Sigma_c} \simeq 0.028 N_{21} v_{rot} / r_{pc}\ ,
\end{equation}
where $N_{21}$ is the neutral hydrogen column density in units
$10^{21}$~cm\mtwo, $v_{rot}$ is the rotational velocity in \kms, and
$r_{pc}$ is the radius in parsecs (based on \citeNP{Toomre64};
\citeNP{Cowie81};
\citeNP{Kennicutt89}). We measure $\Sigma_{HI} / \Sigma_c \sim
0.01$--0.1 for the model \hi\ absorbing disks, based on the columns
reported in Table~\ref{t-diskfits}. The extreme values are $\la 0.001$
for NGC~4151 (AGN disk) and NGC~5506 and 0.45 for NGC~4151 (Galaxy
Disk). 

The \hi\ surface densities generally fall well below the
critical density, but the surface density of molecular and ionized gas
might still push the disks to instability. In the limit where atomic
gas is the minority phase, the stability criterion becomes
\begin{equation}
\frac{\Sigma_{H}}{\Sigma_{HI}} \la
\left(\frac{\Sigma_{HI}}{\Sigma_c}\right)^{-1}\ . 
\end{equation}
That is, for these disks to be unstable to star formation, the
molecular and ionized gas masses would have to be 10--100 times the
atomic gas mass over the inner kiloparsec. 

In global averages, the surface densities of atomic and molecular gas
in spiral galaxies are similar to within factors of a few (e.g.,
\citeNP{Kennicutt90} and references therein). On the other hand,
gas in the centers of spiral galaxies tends to be predominantly
molecular (e.g., \citeNP{JHPB96}; \citeNP{HB97}). Taking NGC~1068 as an example, \citeN{HB95}
estimate a molecular gas column N$_{H_2} \sim 5\times
10^{22}$~cm\mtwo\ near the region we detect \hi\ absorption. The
molecular surface density exceeds the \hi\ surface density by a factor
of $\ga 20$, pushing $\Sigma/\Sigma_c \ga 1$. On the other hand, the
rotation curve remains poorly constrained on this scale; a radial
gradient of the rotation velocity would raise $\Sigma_c$ and stabilize
the disk. 

In NGC~1068, at least, \hi\ is the probably the minority phase within
the inner 100~pc. If other Seyfert galaxies are similarly rich in
molecular gas, the disks traced by \hi\ absorption disks may be
unstable to star formation. Unfortunately, we are unable to test this
suggestion with the present data. Current and future millimeter
surveys, such as that currently taking place at the BIMA telescope
(e.g., \citeNP{HTRSVWBB98}), may better address the stability
issue in other Seyfert galaxies. 

\section{Implications for Unifying Schemes}\label{schemes}

Perhaps the most surprising result of this survey is the lack of \hi\
absorption towards the AGN in Seyfert nuclei, particularly those known
to harbor a heavily obscured central engine. The trivial explanation
is that the central X-ray sources are radio-silent. However, the
well-known hidden Seyfert 1 nuclei are also the most luminous Seyfert
galaxies at radio wavelengths \cite{MHBB92}, and it seems
unlikely that the most luminous radio Seyferts contain radio-silent
AGNs. Another possibility is that the obscuring disks are entirely
molecular, or, alternatively, our sight-line crosses mainly through an
ionized region of the disk. The line-of-sight X-ray columns commonly
exceed $10^{21}$~cm\mtwo\ (Table~\ref{tab-cx}). Regardless of the
viewing geometry, at some distance from the central ionizing source
the obscuring material will become largely neutral. Photodissociation
models predict a substantial atomic column detectable in 21~cm
absorption, even if most of the material is molecular; for example, in
the presence of an Seyfert nucleus, dissociative excitations can
liberate atomic gas within molecular clouds located out to distances of
hundreds of parsecs \cite{MHT96}.

We next consider three ways in which \hi\ absorption from AGN
obscuring disks might be suppressed, namely kinematical line
broadening, free-free absorption of near-nuclear radio continuum, and
excitation effects for the 21~cm line. In each case the implication is
that, ironically, the obscuring disk must be small, probably parsecs
in scale, in order to suppress \hi\ absorption from the obscuring
material. On the other hand, $\ga 100$~pc obscuring disks would not
suffer these suppressing effects, and the prediction would have been
strong \hi\ absorption, contrary to the observations.

\subsection{Kinematical Line Broadening}

For a given mass of obscuring \hi, the absorption line strength is
inversely proportional to the velocity width of the line. In the
environment of a compact mass, the absorption line-width should vary
inversely with the (inner) radius from the AGN, and lines arising from
rapidly moving gas near an AGN may be broadened below
detectability. Without loss of generality, we consider a model where
the ensemble of obscuring gas clouds is supported by random motion
against gravitational collapse. Our 21~cm absorption experiment would
be sensitive to columns:
\begin{equation}
{\rm N_{HI}} \approx 3\times 10^{21}\ {\rm cm^{-2}}\ 
r_{pc}^{-1/2}
M_7^{1/2}
\left(\frac{\sigma_{\rm mJy}}{C_{100}}\right)
\left(\frac{T_{spin}}{8000\ \rm K}\right)
\ ,
\end{equation}
where $r_{pc}$ is the distance from the AGN, $M_7$ is the mass of the
black hole in units $10^7$~\Msun, $\sigma_{\rm mJy}$ is the channel
map sensitivity in units of mJy, and $C_{100}$ is the background
continuum strength in units of 100~mJy (e.g., as is appropriate for
Mrk~348). We have also assumed that the local spin temperature is
enhanced to $\sim 8000$~K, as may obtain in the warm atomic layer near
an AGN \cite{MHT96}. We conclude that the
absorption experiment remains sensitive to column densities that may
be typical for obscured Seyfert 1 nuclei, ${\rm N_H} \ga
10^{23}$~cm\mtwo, in the presence of kinematical line broadening. The
obscuring gas would have to be within $\sim 1\ C_{100}^{-2}$
milliparsec of the AGN to suppress \hi\ absorption below
detectability. 

\subsection{Free-free Absorption}

The hypothesis in this case is that, as has been verified for NGC~1068
(e.g., \citeNP{n1068b}; \citeNP{GBO97};
\citeNP{RCWU98}), the nuclear continuum source has been
attenuated by free-free absorption. Therefore, \hi\ in the neutral
region of the obscuring torus lacks a background source on the
smallest scales, and \hi\ absorption would not be detected. In this
scenario, nuclear radio continuum does not trace emission from the AGN
proper but instead from compact sources in the central, unresolved
region of the radio jet. This is certainly the case for
NGC~1068 (e.g., \citeNP{n1068a}) and Mkn~231 \cite{CWU97}, for example.

Further motivating this hypothesis, the X-ray photoionization models
of \citeN{KL89} and \citeN{NMC94} have demonstrated that X-ray heated plasma within a
parsec-scale obscuring torus can be optically thick to free-free
absorption down to $\sim 1$~cm wavelength.  Observationally, VLBI
surveys are revealing free-free absorbed radio sources in other
Seyfert nuclei (e.g., \citeNP{UWRWFK99}).  \citeN{PFHRD98} find evidence for low-frequency free-free absorption
towards the nucleus of NGC~4151. The paucity of obvious radio cores in
Seyfert nuclei (e.g., \citeNP{Wilson91p227}), even relatively
radio-bright Seyfert nuclei, also argues for free-free suppression of
the Seyfert radio nuclei.

We next demonstrate that the condition of free-free absorption places
limits on the scale of the free-free obscuring plasma, which
in turn limits the inner radius for neutral gas in the obscuring
torus.  For a given free-free opacity (i.e., emission measure), the
free-free radio continuum flux and hydrogen recombination line
flux scale with the projected surface area of the free-free absorbing
plasma. Looking first at radio
continuum emission from the nucleus, the constraint on
the nuclear free-free radio surface brightness gives:
\begin{equation}
r \la 18{\rm\ pc\ } D_{20}\ \sigma^{1/2}_{\rm mJy}\ 
T_4^{-1/2}\ C_f^{-1/2}\ , 
\end{equation}
where $D_{20}$ is the distance in units of 20 Mpc, $\sigma_{\rm mJy}$
is the channel map sensitivity in mJy, $T_4$ is the electron
temperature in units $10^4$~K, and $C_f$ is the covering fraction of
the obscuring medium. This limit also assumes that the brightness
temperature is equal to the electron temperature, appropriate for
optically thick thermal plasma. 

Since, at 1\arcsec\ resolution, any free-free emission from the
obscuring torus would probably be confused with synchrotron emission
from the jet, we are limited not by the continuum sensitivity but
rather the \hi\ absorption line sensitivity. (As new VLBI maps of
Seyfert galaxies are published, continuum constraints on putative
free-free absorbing regions may tighten the constraints
argued here).  X-ray absorbing columns of $10^{23}$~cm\mtwo\ (e.g.,
Table~\ref{tab-cx}) produce an \hi\ opacity of $\tau_{HI} \sim 20
N_{23} / (T_{100}\ \Delta v_{100})$, where $T_{100} = T_S / 100$~K and
$\Delta v_{100}$ is the line-width in units of 100~\kms.  A nuclear
free-free continuum source would be completely attenuated over the
21~cm line profile, and the limit on the AGN radio continuum flux is
therefore the detection limit for a single channel, or roughly 3~mJy
for this experiment. $C_f \sim 50\%$ based on the ratio of Seyfert 1
to Seyfert 2 galaxies \cite{Lawrence91}.  Applying these values,
the obscuring plasma must be smaller than $r \sim 44\ D_{20}$~pc, or
else the background radio continuum provided by thermal free-free
emission would produce an observable \hi\ absorption line.

We next consider the hydrogen recombination line luminosity of the
free-free obscuring region. We choose to consider hydrogen
recombination lines mainly for ease of computation; consideration of
forbidden line emission would require a photoionization model, which
is beyond the scope of this work. Without loss of generality, we
consider only H$\alpha$ line emission. Since, according to unifying
schemes, the ionized region of the torus would not be obscured by dust
in Seyfert~1 galaxies, the spatially-integrated H$\alpha$ luminosities
of the NLR provide an upper limit to the emission from the obscuring
region.  For Seyfert galaxies, the median NLR H$\alpha$ luminosity is
$L_{H\alpha} \sim 10^{41}$~ergs s\mone\ (based on characteristic line
ratios in \citeNP{FO86} and observed H$\beta$
luminosities from \citeNP{Whittle92a}). Taking the usual assumption of
Case-B recombination, the size of the free-free obscuring region must
be:
\begin{equation}
r \la 28{\rm\ pc\ } (L_{H\alpha}/10^{41})\ C_f^{-1/2}\ 
\tau^{-1/2}_{ff}\ T_4^{-0.675}\ ,
\end{equation}
where $\tau_{ff}$ is the free-free opacity at the 21~cm rest wavelength,
and we have used the usual radio approximation for the free-free
opacity law from \citeN{MH67}. 

If free-free suppression of the nuclear continuum source commonly
occurs in Seyfert nuclei, as is at least suggested by this \hi\ 
absorption experiment, the implication is that the obscuring torus
cannot be much larger than a few tens of parsecs. Turning the argument
around, we would not predict substantial free-free absorption from a
100~pc scale obscuring torus; rather, we would have expected 
strong \hi\ absorption against a bright continuum source.

\subsection{Excitation Effects}\label{exciteme}

The third way to suppress \hi\ absorption is to raise the spin
temperature, dropping the opacity for a given foreground column. We
have demonstrated that, for {\em typical} radio source luminosities,
the 21~cm transition in Seyfert obscuring disks is likely to be
thermalized, and \hi\ absorption should be detected (\S\ref{tspin}).
On the other hand, particularly luminous nuclear radio sources will
drive the local spin temperature to the equivalent brightness
temperature of the local 21~cm continuum radiation field (e.g.,
\citeNP{BE69}). The radiation field from a compact
nuclear source will drop as $r^{-2}$, and so excitation effects will
affect 21~cm absorption only if the obscuring gas is located near the
central engine. Since the hypothesis is that we see the bright radio
nucleus directly, this case can be viewed as the opposite extreme of
the free-free absorption scenario presented above.

In particular, spin temperature effects may explain the lack of \hi\ 
absorption towards Mrk~348. Mrk~348 has an unusually bright nuclear
radio source, $S_{\nu} \approx 100$--500~mJy at 21~cm. This source is
variable on time-scales of years, marking it as a likely location of the
AGN (e.g., \citeNP{NdB83}; \citeNP{UPNdB84}). Assuming the source is much smaller than 1~pc, the 21~cm
continuum temperature as a function of radius is
\begin{equation}
T_R \approx 2.4\times 10^{10}\ {\rm K}\ r_{pc}^{-2}\ .
\end{equation}
Applying the formalism of \citeN{BE69}, the 21~cm
radiation field will dominate the effects of particle collisions at
radii:
\begin{equation}
r \la 5{\rm\ pc}\ S_{100}^{1/2}\ n_5^{-1/2}\ D_{60}\ T_{100}^{0.1}\ ,
\label{eq-excite}
\end{equation}
where $S_{100}$ is the continuum source brightness in units of
100~mJy, $n_5$ is the gas density in units $10^5$~cm\mthree, $D_{60}$
is the distance to the source in units of 60~Mpc, and $T_{100}$ is the
kinetic temperature in units of 100~K. The normalizations are
specific to Mrk~348, but the result can be scaled and applied to any
compact radio source. We estimate that the spin temperature might
be as high as $\sim 10^6$~K at a radius of 1~pc, and, as a result,
21~cm absorption opacity would be suppressed by a factor of $10^4\ 
T_{100}^{-1}$!

\subsection{Conclusions}

\citeN{SPNSM89} originally proposed that the obscuring
medium in Seyfert galaxies might span scales much larger than the
conventional torus model.  More recently, \citeN{MGT98} reported that optical images of Seyfert 2 galaxies are
more likely to show $\sim 100$~pc dust absorption features than
optical images of Seyfert~1 galaxies. They argued that these galactic
dust lanes are primarily responsible for hiding the central engine and
BLR in Seyfert~2 nuclei. The characteristic scale of the obscuring
medium in this galactic dust model is $\ga 100$~pc. To some degree
galaxy disks must obscure Seyfert nuclei, since Seyfert nuclei are
rarely found in edge-on disk galaxies (\citeNP{Keel80};
\citeNP{SMSE97}).

To explain the surprising lack of \hi\ absorption towards known,
hidden Seyfert 1 nuclei requires, ironically, obscuring disks on 10~pc
scales or less. Turning the arguments around, we would expect that
kinematical line broadening, free-free absorption, and excitation
effects would have little effect on 21~cm absorption from gas on
100~pc scales. Therefore, the galactic dust model 
predicts pronounced \hi\ absorption towards radio-bright Seyfert
nuclei. We find instead that the dust lanes observed by \citeN{MGT98} are probably the source of the off-nuclear \hi\ 
absorption reported here. We conclude that the inner scale of the
obscuring medium in hidden Seyfert 1 galaxies must be less than a few
tens of parsecs, in better agreement with the classical parsec-scale torus
scenario.

It is all but certain that free-free suppression of the nuclear
continuum source has precluded the detection of 21~cm absorption
towards NGC~1068. It seems very likely that excitation effects have
suppressed 21~cm absorption in Mrk~348. The case of Mrk~3 is more
difficult to address. \citeN{KGPS99} identify a compact,
flat spectrum VLBI source as the nucleus. The source is weak, only
9~mJy at 18~cm, and so excitation effects will be important only on
sub-parsec scales (scaling from Eqn~\ref{eq-excite}, above). On the
other hand, this source might be intrinsically steep-spectrum but
free-free absorbed at 18~cm, in which case it need not be identified
as the nucleus. Additional spectral information is needed to interpret
the nuclear radio source and the non-detection of 21~cm absorption.
Otherwise, we have demonstrated that for plausible conditions in
Seyfert nuclei, either free-free absorption or excitation effects
would suppress 21~cm absorption from parsec-scale obscuring disks.
These effects might naturally explain the tendency for 21~cm
absorption to avoid the nuclei of {\em obscured} Seyfert galaxies.

\section{Implications for AGN Fueling}~\label{fueling}

One of the long-standing problems of AGN physics is the nature of the
fuel source for the central engine. Based, for example, on source
counts, or the sizes of radio jets, the lifetime of an AGN is expected
to exceed $\sim 10^8$~yr (e.g., as reviewed by
\citeNP{Peterson97}). Supporting a typical Seyfert luminosity of
$10^{44}$~ergs s\mone\ requires a fueling rate of about $0.02\
(\epsilon/0.1)$~\Msun\ yr\mone. Integrated over its lifetime, the
central engine needs a total mass of $\sim 2\times 10^6\
(\epsilon/0.1)$~\Msun. This mass is comparable to the central \hi\
masses derived in this absorption experiment (Table~\ref{t-diskfits}).
The required infall velocities are low over the inner few hundred
parsecs, only a few \kms, and the
\hi\ kinematics are compatible with our limits on the streaming
velocities ($v_{rad} \sim$ few tens of \kms; \S\ref{streaming}).  We
may also conclude that the inferred \hi\ mass of the central galactic
disks of Seyfert galaxies is sufficient to power the central engine
over its lifetime.

Bringing the gas to the central engine poses a more serious problem.
The main obstacle is removing sufficient angular momentum to bring gas
down to the inner parsec or so, where viscous dissipation takes over
(e.g., as reviewed by \citeNP{Phinney94p1}). Stellar bars provide
gravitational torque and may promote cloud-cloud collisions, funneling
gas down at least as far as an inner orbital resonance.
\citeN{SFB89} proposed that mass
accumulation at resonances can trigger the formation of nested bars.
In principle, bars might nest down to arbitrarily small scales,
carrying gas down to correspondingly small volumes.

A basic test of this scheme is to compare the frequency of bars among
active and non-active spiral galaxies. Taking advantage of new
near-infrared imaging arrays, recent surveys have found that, in
general, spiral galaxies are frequently barred ($\sim 70\%$), but the
incidences of bars among active and non-active spiral galaxies are
comparable \cite{MR97}. Moreover, the presence or
strength of a bar appears to have little or no impact on the strength
of the nuclear activity \cite{HFS97}.  As mass
accumulates at the center of a bar, stellar orbits in the bar are
perturbed, leading to the destruction of the bar (e.g.,
\citeNP{FB93}; \citeNP{NSH96}).
This mechanism might explain the absence of bars in a fraction of
Seyfert nuclei. Nevertheless, the results of the imaging surveys raise
two questions for the bar fueling scenario. First, there is growing
evidence for black holes in the centers of both active and non-active
galaxies (e.g., \citeNP{KR95};
\citeNP{Miyoshi95}; \citeNP{maserpaper}; \citeNP{EG96}). Why are the black holes in barred, non-active spirals
dormant? Second, other than the presence of a bar, what parameter
might distinguish the host galaxies of active and non-active spirals?

Our study suggests that the gas mass of the central galactic disk may
be an important criterion distinguishing active and non-active
spirals. The high detection rate of \hi\ absorption argues that the
central, 100~pc scale disks of Seyfert nuclei are gas-rich, with
azimuthally averaged surface densities $\Sigma_{HI}({\rm Seyfert}) \ga
6\ (C_f[{\sc Hi}]/0.7)$~\Msun\ pc\mtwo, where $C_f[{\sc Hi}]$ is the
areal covering fraction of \hi\ over the central galactic disk,
normalized to the detection rate.  Such
high surface densities should be easily detected in 21~cm emission in
normal spirals.  In contrast, non-active spiral galaxies commonly show
an \hi\ depression over the inner few {\em kiloparsecs}
(e.g., \citeNP{Roberts75p309}; \citeNP{vW+83};
\citeNP{GH88p522}), with column densities dropping
below detections limits of $\sim 1$~\Msun\ pc\mtwo. 

It is difficult to address this issue further, because there has been
no systematic study of central \hi\ depressions in spiral galaxies.
The reason is that, except in the dozen or so nearest galaxies,
current aperture synthesis telescopes are generally not sensitive to
\hi\ emission at sub-kpc resolution. To illustrate, we attempted to
generate a comparison sample based on published \hi\ maps. The
selection criteria were (1) Hubble type earlier than Sb, since most
Seyfert galaxies reside in early type spiral hosts; and (2) linear
resolution better than 1~kpc, in order to reduce confusion with the
outer disk. We chose galaxies from the Revised Shapley-Ames catalog
\cite{ST81} and referenced them against the Martin's
(1998) \nocite{Martin98} bibliographic index of 21~cm maps of
galaxies. This selection process picked out 18~candidates for the
comparison sample, but all but two of the candidates were either
edge-on, active (Seyfert, starburst, or LINER), or strongly
interacting with a close companion.  This result reflects an
observational bias towards the unusual but precludes an meaningful
comparison between active and non-active galaxies.

Nevertheless, we emphasize that feeding the AGN may require not only a
stellar bar, but also an excess of gas mass in the inner kiloparsec as
a source of fuel. Numerical studies of gas dynamics find that stellar
bars remove a steady fraction $f_{\rm bar}$ of the local gas mass per
rotation; $0.01 \la f_{\rm bar} \la 0.1$ (\citeNP{Athanassoula92};
\citeNP{WH92}; \citeNP{FB93};
\citeNP{HS94}; \citeNP{Athanassoula94p143}). The gas
mass required to fuel an AGN through bar-feeding is 
\begin{equation}
M = 1.01\times 10^7\ M_{\odot}\ L_{44}
                                \left( \frac{\epsilon}{0.1} \right)^{-1}
                                \left( \frac{f_{\rm bar}}{0.1} \right)^{-1}
                                \Omega_{100}^{-1}\ ,
\end{equation}
where $L_{44}$ is the luminosity of the AGN in units $10^{44}$~ergs
s\mone, and $\Omega_{100}$ is the bar pattern speed in units 100~\kms\ 
kpc\mone. Accepting the considerable uncertainties in the
efficiencies, the observed \hi\ absorption masses in active spirals,
which give a lower limit to the total gas mass over the central disk,
are comparable to the numerical predictions for bar feeding. Otherwise
having a stellar bar and a central, supermassive black hole,
non-active spirals may simply lack a central fuel source to power an
AGN or nuclear starburst. Testing this scenario will require new \hi\ 
emission observations of a sample of active and non-active spiral
galaxies, reasonably matched in Hubble type and physical resolution
(distance).  The prediction is that, over the inner kiloparsec, the
mean \hi\ surface density of Seyfert galaxies will exceed the \hi\ 
surface density of non-active spirals.

\section{Conclusions} \label{conclusions}

Our main conclusions are as follows:

\begin{enumerate}
\item \hi\ 21~cm absorption in Seyfert and starburst galaxies appear
  to trace gas in the central disks of the host galaxy. The radial scale
  of the central absorbing disks is hundreds of parsecs (i.e., less
  than a kiloparsec). High-velocity \hi\ absorption, Doppler-shifted
  away from the systemic velocity of the host galaxy, can usually be
  explained by normal rotational motion rather than radial streaming
  (e.g., NGC~2110). 

\item The \hi\ kinematics are compatible with, but do not require,
  streaming velocities of several tens of \kms. The corresponding mass
  inflow rates can be several \Msun\ yr\mone, generally exceeding the
  mass transfer rate required to fuel a Seyfert nucleus.

\item NGC~4151 appears to be a special case \cite{MHPMKA95}. The kinematics and distribution of the \hi\ absorption are
  consistent with a compact, parsec-scale disk oriented perpendicular
  to the radio jet. In contrast, the \hi\ properties are not
  compatible with the larger scale galaxy disk orientation and
  rotation sense.

\item \hi\ absorption in Seyfert galaxies generally avoids the nucleus
  but is more commonly seen against an extended radio component. The
  21~cm \hi\ columns are not correlated with the hydrogen column
  derived from X-ray spectroscopy, even in known hidden Seyfert~1
  nuclei. We propose that Seyfert nuclei are commonly free-free
  absorbed, and no 21~cm continuum from the AGN escapes to the neutral
  layer of the obscuring medium. Alternatively, bright nuclear
  continuum sources, such as in Mrk~348, can enhance the spin
  temperature and suppress 21~cm line absorption. These suppression
  mechanisms constrain the size of the obscuring medium to smaller
  than a few tens of parsecs.

\item We suggest that the central disks of active spiral galaxies may
  show higher \hi\ surface densities compared with non-active spirals.
  This result may resolve the observational challenges for the bar
  fueling model of AGNs. That is, to be active a spiral galaxy might
  require both a central stellar bar {\em and} an excess of neutral
  gas. Testing this prediction will require further studies of 21~cm
  emission from matched samples of active and non-active spirals. It
  is not clear from this study whether we are witnessing gas en route
  from the outer galaxy (i.e., the fuel line) or a local reservoir from
  which the nucleus derives its fuel (i.e., the fuel tank). 
\end{enumerate}

\acknowledgements{
We would like to acknowledge helpful conversations with Andrew Baker,
Mort Roberts, Andrew Wilson, and Jim Ulvestad regarding clarifications
and the conclusions of this work. We thank the staff at the VLA Array
Operations Center for technical assistance with the observations and
data reduction. JFG received support from the STScI Collaborative
Visitor's program and a Jansky Fellowship at NRAO, Charlottesville.
}

\appendix

\section{Comments on the Individual Sources}\label{individuals}

\subsection{Detected Sources}

\noindent{\em Mkn 6} is a Seyfert~1.5 galaxy. \hi\ absorption was
originally reported by \citeN{HBS78} and
\citeN{BH83}, whose detections we confirm. We have
reported a more detailed follow-up MERLIN observation \cite{GHPM98}. The nuclear radio source, marginally resolved by the present
observation, collimates into a 1\arcsec\ (360~pc) radio jet
\cite{Kukula96}. \hi\ absorption is detected against the northernmost
radio component of the 1\arcsec\ nuclear jet. The offset absorption
appears to arise from a hundred-pc scale arm or dust lane crossing
just north of the AGN. The AGN itself may be partially absorbed by the
dust lane, which may lend to its classification as a Seyfert
1.5. Otherwise, there are no radio candidates for the AGN. Recent ASCA
spectroscopy suggests a high emission measure to the nucleus, and so
the radio nucleus may be free-free absorbed \cite{FBEFIM99}.

\noindent{\em Mkn 231} is a luminous and morphologically peculiar
Seyfert 1 galaxy. \hi\ absorption was first reported by
\citeN{HBS78} and \cite{vdHHD82}. Our
VLA spectra confirm these earlier detections. The VLBA observations of
\citeN{CWU97} resolve the \hi\ absorption
against a 0\farcs44 ($\sim 350$~pc) radio continuum disk surrounding
the AGN. The \hi\ absorption displays a $\sim 220$~\kms\ velocity
gradient along the major axis of the diffuse radio continuum disk, and
\citeN{CWU97} argue that the \hi\ absorption
arises in the atomic component of a molecular disk mapped by
millimeter-wave CO emission \cite{BS96}.

\noindent{\em NGC 1068} is the prototype of Seyfert 2 nuclei but is
also the first Seyfert found to have a Seyfert 1 spectrum in polarized
light \cite{AM85}. \citeN{BH83} originally reported
\hi\ absorption seen as a sharply defined trough in the \hi\ emission
profile. Our observations resolve out \hi\ emission, and the \hi\
absorption is spatially resolved. The main results were originally
presented in \citeN{GBOBP94}. The bulk of the absorption
traces the inner disk of the host galaxy, seen against the
southwestern radio structure. There is kinematic evidence that the gas
is responding to the kiloparsec scale stellar bar. Closer to the
nucleus the absorption lines broaden, and there appear multiple
velocity components. The opacity of the absorption line gas also
appears to increase, but towards the nucleus the opacity seems to drop
sharply. VLBI studies have failed to detect the nucleus (radio
component S1) in 21 cm radio continuum \cite{RCWU98}. The
apparent drop in opacity is probably a resolution effect owing to the
lack of a background radio source. There may be a broad absorption
profile very near the nucleus, whose kinematics are consistent with
the rotating \h2o\ maser disk \cite{maserpaper}, but bandpass
calibration uncertainties preclude a confident detection in the
present observations.

\noindent{\em NGC 2110} is an X-ray bright Seyfert 2 galaxy
\cite{MRCVvP79}.  At sub-arcsecond resolution, the
``figure-eight'' radio structure of our 21~cm continuum image resolves
into a lazy-{\bf S}-shaped radio jet \cite{WU83}. We detect redshifted \hi\ absorption offset to the south
towards the weaker radio lobe.

\noindent{\em NGC 2992} is a narrow line X-ray galaxy \cite{WPBT80}. The radio structure comprises an unresolved nuclear source
surrounded by figure-8 loops \cite{WM88}. \hi\
absorption is detected offset from the radio nucleus against one of
the extended lobes. The host galaxy is viewed nearly edge-on, and so
it seems likely that the \hi\ absorption traces gas in the foreground
disk. 

\noindent{\em NGC 3079} is classified as a LINER (\citeNP{Heckman80}) with evidence for active star-formation (e.g., \citeNP{CB88}; \citeNP{IS90}). The \hi\ absorption
was discussed briefly by \citeN{GBOBP94} but in more
detail by \citeN{BI95}. MERLIN observations well resolve
the rotational trend of the absorption line gas \cite{PMGBO96}, which matches the sense of the rotation traced by \hi\
emission \cite{IS91}. The bulk of the absorption
probably arises from the edge-on but normally rotating disk, observed
in silhouette against extended, diffuse radio emission associated with
star-formation \cite{PMGBO96}. Based on new VLBI
observations, \citeN{SIN98p219} report that the \hi\
absorption viewed towards the central VLBI radio jet appears to be
counter-rotating relative to the gas motions in the outer galaxy, but
it is unclear whether the \hi\ absorption on VLBI scales traces gas in
rotation as opposed to, say, nuclear-driven outflow. The apparent
complexity of the broad absorption line profile may reflect the
structure of the VLBI jet as much as the kinematics and distribution
of the absorbing gas.

\noindent{\em NGC 3504} is a barred starburst nucleus originally
included on S\'ersic's list of peculiar nuclei \cite{Sersic73}. The
nuclear radio emission comprises a compact, unresolved core surrounded
by a $\sim 10$\arcsec\ diffuse halo (e.g., \citeNP{SPUA94}). The halo probably traces synchrotron emission from a
12\arcsec\ diameter starburst ring, revealed in near-infrared color
maps \cite{ECSM97}. \citeN{Dickey86} reported a
tentative detection of \hi\ absorption. We clearly detect an
absorption line towards the unresolved radio nucleus of this galaxy.

\noindent{\em NGC 4151} is a well-studied Seyfert 1.5 galaxy. \hi\
absorption was originally detected in this experiment and followed-up
by MERLIN observations \cite{MHPMKA95}. The nuclear radio
structure resolves into a $\sim 5\arcsec$ ($\sim 320$~pc) radio jet
(\citeNP{PKLMABOU93} and references therein). \hi\ absorption
is detected only towards the radio component C4 \cite{MHPMKA95}, which itself resolves into a double source on VLBI
baselines \cite{HPUBGP86}. The tight confinement of the
\hi\ absorption to the VLBI portion of the radio jet argues that the
absorption arises from near-nuclear gas, perhaps a disk surrounding
the central engine, at tens of parsecs from the central engine
\cite{MHPMKA95}.

\noindent{\em NGC 5506} is an X-ray bright Seyfert 2 galaxy
(\citeNP{WPFB76}; \citeNP{Rubin78}).  \hi\ absorption
was originally reported by \citeN{TW82}. The radio
emission is dominated by a compact radio source, unresolved at $\sim
0\farcs1$ resolution \cite{WM87}. A low surface
brightness arc extends $\sim 0\farcs5$ north from the nucleus
(\citeNP{UPBH86}; \citeNP{WM87}), and faint
emission extends $\sim 30\arcsec$ from the nucleus \cite{CBGOC96}. The host galaxy is viewed nearly edge-on, and it seems
likely that most of the absorption arises in foreground disk gas.

\subsection{Non-detections}

\noindent{\em Mkn 3} is a Seyfert 2 nucleus harboring a hidden BLR
\cite{MG90}. The radio structure is linear,
resolving into a 2\arcsec\ ($\sim 600$~pc) jet \cite{Kukula93}. Surprisingly, we do not detect \hi\ absorption in this
hidden Seyfert 1. On the other hand, there are no flat spectrum
components that clearly mark the location of the central engine
\cite{Kukula93}. It may be that the nuclear radio source is free-free
suppressed or radio silent, and so there is no background radio source
to illuminate the material obscuring the central engine.

\noindent{\em Mkn 348 (NGC 262)} is a Seyfert 2 galaxy that is unusual
in two respects. Firstly, it has a very large \hi\ disk grown perhaps
by a tidal encounter with a neighboring spiral (\citeNP{HSBS82}; \citeNP{SvGHS87}). Secondly, there is evidence
for a hidden BLR in this source (e.g., \citeNP{MG90}. Although the nucleus hosts a very bright subarcsecond radio
triple (e.g., \citeNP{NdB83}), we detect no \hi\
absorption in this experiment. \citeN{HUR97}
report evidence for free-free absorption below 6~cm wavelength, but it
is unclear which components of the radio triple might be
absorbed. (This result is consistent with the spectral decomposition
of \citeNP{NdB83}). The lack of \hi\ absorption from
the medium obscuring the BLR might again be ascribed to free-free
suppression of the nuclear radio continuum source at 21~cm wavelength.

\noindent{\em Mkn 668 (OQ +208)} has the optical spectrum of a
Seyfert~1 but, given its radio power, may be more appropriately
classified as a broad line radio galaxy (e.g., \citeNP{SBDOBFF97} and references therein). Mkn~668 is also notable for being
one of the nearest GHz-peaked spectrum radio sources \cite{OBS91}. VLBI observations resolve the compact
radio source into a 7~mas ($\sim 5$~pc) double \cite{SBDOBFF97}.

\noindent{\em NGC 2639} is a LINER hosting an \h2o\ megamaser
\cite{WBH95}. As may also be the case for the
megamaser sources NGC~1068 and NGC 4258, free-free absorption may
suppress 21~cm continuum within the maser disk (\citeNP{GBO97};
\citeNP{HGMDINM98}). This may explain the lack of \hi\ absorption
corresponding to the \h2o\ maser. Alternatively, there may be very
little atomic gas associated with the \h2o\ masers, but that seems
unlikely in the photodissociative environment of an AGN. The radio
emission resolves into a 1\farcs4 linear structure \cite{UW89}, but,
mainly because no spectral index information is available, there is
presently no radio candidate for the AGN.

\makeatletter
\def\jnl@aj{AJ}
\ifx\revtex@jnl\jnl@aj\let\tablebreak=\nl\fi
\makeatother
\singlespace

\begin{figure}
\caption{Global (spatially-integrated) \protect\hi\ absorption spectra
  of Seyfert 1 nuclei. The spectra are continuum-subtracted and
  corrected back to zero rest velocity based on the observed
  recessional velocities in Table~\protect\ref{sourceprops}. The only
  non-detection among the Seyfert 1 nuclei was
  Mkn~668.\label{sy1spec}}
\end{figure}

\begin{figure}
\caption{Global \protect\hi\ absorption spectra of Seyfert 2 and
  narrow-line X-ray nuclei. The spectra are continuum-subtracted and
  corrected back to zero rest velocity based on the observed
  recessional velocities in Table~\protect\ref{sourceprops}. The
  non-detections are Mkn~3 and Mkn~348. \label{sy2spec}}
\end{figure}

\begin{figure}
\caption{Global \protect\hi\ absorption spectra of starburst and LINER
  nuclei. The spectra are continuum-subtracted and corrected back to
  zero rest velocity based on the observed recessional velocities in
  Table~\protect\ref{sourceprops}. The only non-detection is NGC~2639.
  \label{sbspec} }
\end{figure}

\begin{figure}
\caption{Overlay of integrated \protect \hi\ absorption (contours) on
  radio continuum (inverse grayscale) of NGC 1068. The contour levels are
  0.21, $-0.21$, $-$0.41, $-$0.60, \& $-$0.80 Jy~\kms. The maximum grayscale
  level is clipped at 0.3~Jy. The star marks the location of the
  nucleus. The beam is indicated by the filled
  ellipse in the lower left corner.\label{n1068tot}}
\end{figure}

\begin{figure}
\caption{Channel-map overlays of \protect \hi\ absorption (contours) on
  radio continuum (inverse grayscale) of NGC 1068. The contour levels
  are $-13.9$, $-9.2$, $-6.1$, $-4.1$, \& $-2.7$ ($3\sigma$)
  mJy~beam\mone. The grayscale stretch is logarithmic, and the maximum
  grayscale level displayed is clipped at 1.0~Jy. The channel
  velocities, in \kms, are provided in the upper right corner of each
  panel. \label{n1068panel}}
\end{figure}

\begin{figure}
\caption{Overlay of integrated \protect \hi\ absorption (contours) on
  radio continuum (inverse grayscale) of NGC 2110. The contour levels
  are 0.45, $-0.45$, $-$0.57, $-$0.71, \& $-$0.90 Jy~\kms. The star
  marks the location of the radio nucleus. The beam is indicated by
  the filled ellipse in the lower left corner. \label{n2110tot}}
\end{figure}

\begin{figure}
\caption{Overlay of integrated \protect \hi\ absorption (contours) on
  radio continuum (inverse grayscale) of NGC 2992. The contour levels
  are 0.36, $-0.36$, $-$0.49, $-$0.66, \& $-$0.90 Jy~\kms. The star
  marks the location of the radio nucleus. The beam is indicated by
  the filled ellipse in the lower left corner. \label{n2992tot}}
\end{figure}

\begin{figure}
\caption{Overlay of integrated \protect \hi\ absorption (contours) on
  radio continuum (inverse grayscale) of NGC 3079. The contour levels
  are 0.57, $-0.57$, $-$1.1, $-$2.2, $-4.3$ \& $-$8.4 Jy~\kms. The
  continuum is displayed with a logarithmic stretch. The star marks
  the peak of the radio continuum, which is presumably the nucleus.
  The beam is indicated by the filled ellipse in the lower left
  corner. \label{n3079tot}}
\end{figure}

\begin{figure}
\caption{Channel-maps of the \protect\hi\ absorption (contours)
  towards NGC 3079. The contour levels are $-40.5$, $-19.1$, $-9.0$,
  $-4.2$, $-2$, \& $-2$ mJy~beam\mone. The continuum is displayed as
  contours on greyscale in the lower left panel. The star symbol marks
  the location of the radio continuum peak on the channel maps. The
  continuum contour levels are $-0.30$, 0.30, 0.66, 1.5, 3.2, 7.0,
  15.4, 34.0, \& 74.7~mJy~beam\mone. The channel velocities, in \kms,
  are provided in the upper right corner of each panel. A $\sim
  0\farcs3$ scale velocity gradient, with velocities increasing from
  northwest through southeast, is marginally resolved by our data
  (compare with Pedlar et al. 1996).\label{n3079panel}}
\end{figure}

\begin{figure}
\caption{Overlay of integrated \protect \hi\ absorption (contours) on
  radio continuum (inverse grayscale) of NGC 4151. The contour levels
  are 0.3, $-0.3$, $-$0.45, $-$0.70, \& $-$1.1 Jy~\kms. The beam is
  indicated by the filled ellipse in the lower left corner.
  \label{n4151tot}}
\end{figure}

\nocite{PMGBO96}

\begin{figure}
\caption{Overlay of integrated \protect \hi\ absorption (contours) on
  radio continuum (inverse grayscale) of Mrk~6. The contour levels are
  0.19, $-0.19$, $-$0.46, $-$1.1, \& $-$2.7 Jy~\kms. The star marks
  the location of the optical nucleus (Clements 1981).
  The beam is indicated by the filled ellipse in the lower left
  corner.\label{mk6tot}}
\end{figure}

\nocite{Clements81}

\begin{figure}
\caption{Plots of $T_S$(21~cm)$/T_K$ for varying kinetic temperature
$T_K$ as a function of neutral hydrogen density
  $n_H$. The models plotted in the bottom panel include only the
  effects of 21~cm absorption and collisional excitation, but the
  plots in the top panel include the effects of Ly-$\alpha$ pumping of
  the ground state of \hi. The three curves on each panel correspond
  to $T_K = $ 3, 10, and 100~K. We modeled the radio and UV continuum
  properties after NGC~4151, assuming a distance of 50~pc between the
  AGN and absorbing cloud (further details are provided in the
  text). Irrespective of the presence of a UV pump, the 21~cm
  transition thermalizes for $n_H \ga 1000$~cm\mthree. Note that
  21~cm continuum excitation may affect $T_{spin}$ in Mrk~348 and
  other sources with bright, flat-spectrum cores (see
\S\protect\ref{exciteme}). 
\label{fig-tspin} }
\end{figure}

\begin{figure}
\caption{Absorbing \hi\ column density plotted vs. galaxy
  inclination. The columns for Seyfert galaxies are plotted as filled
  circles; upper limits are indicated by arrows. Measurements from
  Dickey (1986) have been included. The Seyfert columns assume a
  characteristic $T_{spin} = 100$~K, which may be a lower limit to the
  true spin temperature. The dashed line traces the secant curve
  appropriate for a face-on column density of $2\times
  10^{21}$~cm\mtwo, representative of the Seyfert detections. The open
  circles mark Galactic absorption data from Payne, Salpeter, \&
  Terzian (1982) and Schwarz, Ekers, \& Goss (1982). The Galaxy
  absorption columns have been corrected for measured $T_{spin}$. The
  effective inclination of the Galaxy absorption measurements was
  taken to be the complement of the Galactic latitude. The mean
  face-on column for the Galactic absorption measurements is $3\times
  10^{20}$~cm\mtwo\ and is traced by the solid secant curve. Compare
  with Figure~5 of Dickey (1986).
\label{nh-vs-inc} }
\end{figure}

\nocite{PST82}
\nocite{SEG82}
\nocite{Dickey86}

\begin{figure}\singlespace
\caption{A plot comparing \hi\ columns measured by 21~cm absorption
  (x-axis) with total hydrogen columns measured by model fits to
  soft X-ray spectra (y-axis). We assumed $T_{spin} = 100$~K in
deriving $N_{HI}$; the $N_{HI}$ measurements are actually lower
limits. Narrow-line objects, including Seyfert 
  type 2, LINER, and starburst nuclei, are marked by filled squares,
  and broad-line objects are marked by unfilled squares. The line
  traces $N_{HI}({\rm 21\ cm}) = N_{H}$(X-ray).
\label{fig-c21cx} }
\end{figure}

\clearpage

\begin{figure}
\caption{A kinematic model fit for the \hi\ absorbing disk of
  Mkn~6. The data are MERLIN observations from Gallimore et al.
  (1998). {\em Left:} A schematic of the \hi\ absorbing disk model.
  The disk has been shaded so that the far side appears lighter than
  the near side, as if it were viewed through an atmosphere; the
effect is to mimic the appearance of a dust lane. The disk shading of the
is lightened where there is significant background
continuum, mainly to improve the visibility of the continuum
contours. We erased the disk shading where the radio continuum
structure lies foreground to the disk to illustrate the relative
orientation of the disk and jet. The scale
  height of the disk has been exaggerated for presentation. The
  contours trace the 20~cm radio continuum image. Note that the outer
  radius of the \hi\ disk is a lower limit set by the extent of the
  observed \hi\ absorption. {\em Right: } Model spectra. The top panel
  shows the fit (thick line) against the \hi\ spectrum (thin line), and the
  bottom panel shows the residuals.
  \label{mkn6norm} }
\end{figure}

\begin{figure}
\caption{A kinematic model fit for the \hi\ absorbing disk of
  NGC~1068. {\em Left:} A schematic of the \hi\ absorbing disk model,
  following the convention of Fig.~\protect\ref{mkn6norm}. The
  contours trace the 20~cm radio continuum image. The outer
  radius of the \hi\ disk is a lower limit set by the extent of the
  observed \hi\ absorption. {\em Right: } Model spectra. The top panel
  shows the fit (thick line) against the \hi\ spectrum (thin line), and the
  bottom panel shows the residuals.
  \label{n1068norm} }
\end{figure}

\begin{figure}
\caption{A kinematic model fit for the \hi\ absorbing disk of
  NGC~2110. {\em Left:} Schematic view, following the convention of
  Fig.~\protect\ref{mkn6norm}. The contours trace a 6~cm radio
  continuum image from Ulvestad \& Wilson (1984). We display the
  higher resolution VLA image to emphasize the \hi\ absorbed radio
  structure. The outer radius of the \hi\ disk is a lower limit set by
  the extent of the observed \hi\ absorption. {\em Right:} Model
  spectra. The top panel shows the fit (thick line) against the \hi\ spectrum
  (thin line), and the bottom panel shows the residuals.
  \label{n2110norm} }
\end{figure}

\nocite{UW84b}

\begin{figure}
\caption{A kinematic model fit for the \hi\ absorbing disk of
  NGC~2992. {\em Left:} Schematic view, following the convention of
  Fig.~\protect\ref{mkn6norm}. The contours trace a 6~cm
  radio continuum image from Ulvestad \& Wilson (1984). The outer
  radius of the \hi\ disk is a lower limit set by the extent of the
  observed \hi\ absorption. {\em Right: } Model spectra. The top panel
  shows the fit (thick line) against the \hi\ spectrum (thin line), and the
  bottom panel shows the residuals. \label{n2992norm} }
\end{figure}

\begin{figure}
\caption{A kinematic model fit for the \hi\ absorbing disk of
  NGC~3079. {\em Left:} Schematic view, following the convention of
  Fig.~\protect\ref{mkn6norm}. The contours are 20~cm radio
  continuum. {\em Right: } Model spectra. The top panel shows the fit
  (thick line) against the \hi\ spectrum (thin line), and the bottom panel
  shows the residuals. \label{n3079norm} }
\end{figure}

\begin{figure}
\caption{A kinematic model fit for the \hi\ absorbing disk of
  NGC~3504. {\em Left:} Schematic view, following the convention of
  Fig.~\protect\ref{mkn6norm}. The outer radius of the disk is poorly
constrained, because the background continuum is so weak, and
therefore offset \hi\ absorption contributes little to the global
profile. The contours are 20~cm radio 
  continuum. {\em Right: } Model spectra. The top panel shows the fit
  (thick line) 
  against the \hi\ spectrum (thin line), and the bottom panel shows the
  residuals. \label{n3504norm} }
\end{figure}

\begin{figure}
\caption{A kinematic model fit for the \hi\ absorbing disk of
  NGC~4151. {\em Left:} Schematic view, following the convention of
  Fig.~\protect\ref{mkn6norm}. The contours are 20~cm radio
  continuum modeled after the VLBI observations of Ulvestad et
  al. (1998). In this model, the disk is oriented with 
  the same inclination and position angle as the outer galaxy
  disk. {\em Right: } The top panel shows the fit (thick line) 
  against the \hi\ spectrum (thin line), and the bottom panel shows the
  residuals. Note that this model requires a counter-rotating
  disk. \label{n4151disknorm} } 
\end{figure}

\nocite{URCW98}

\begin{figure}
\caption{A kinematic model fit for the \hi\ absorbing disk of
  NGC~4151. {\em Left:} Schematic view, following the convention of
  Fig.~\protect\ref{mkn6norm}. The contours are 20~cm radio continuum
  modeled after the VLBI observations of Ulvestad et al. (1998). In
  this model, the disk is oriented with the same inclination and
  position angle as that inferred for an AGN disk (Mundell et al.
  1995). {\em Right: } The top panel shows the fit (thick line)
  against the \hi\ spectrum (thin line), and the bottom panel shows the
  residuals. \label{n4151nucnorm} }
\end{figure}

\begin{figure}
\caption{A kinematic model fit for the \hi\ absorbing disk of
  NGC~5506. {\em Left:} Schematic view, following the convention of
  Fig.~\protect\ref{mkn6norm}. The contours are 20~cm radio
  continuum. {\em Right: } Model spectra. The top panel shows the fit
  (thick line) 
  against the \hi\ spectrum (thin line), and the bottom panel shows the
  residuals. \label{n5506norm} }
\end{figure}

\begin{deluxetable}{lrrrrrrrrrll}
\tablewidth{0pt}
\tablecaption{Source Properties}
\tablehead{
\mc{Source} & \mccc{Right Ascension} & \mccc{Declination} & \mcc{$cz$} & \mc{Ref.} & \mc{Activity} & \mc{Morphology}\\
 & \mccc{(1950)} & \mccc{(1950)} & \mcc{\kms} & & & \mc{(NED)\tablenotemark{a}}}
\startdata
Mkn 348  & 00 & 46 & 04.85 & $+31$ & 41 & 04.5 & 4507 & $\pm 4$ & 3 & 2(1)  & SA(s)0/a\\
NGC 1068 & 02 & 40 & 07.07 & $-00$ & 13 & 31.2 & 1150 & $5$ 
                     & 1 & 2(1)  & SA(rs)b\\
NGC 2110 & 05 & 49 & 46.38 & $-07$& 28 & 00.9 & 2335 & 20 & 2 & 2     & SAB0$-$  \\
Mkn 3    & 06 & 09 & 48.24 & $+71$ & 03 & 10.5 & 3998 &  6 & 3 & 2(1)  & S0\\
Mkn 6    & 06 & 45 & 43.96 & $+74$ & 29 & 09.9 & 5640 & 10 & 7 & 1.5   & SAB0+\\
NGC 2639 & 08 & 40 & 03.12 & $+50$ & 23 & 10.2 & 3336 & 11 & 3 & L     & SA(r)a \\
NGC 2992 & 09 & 43 & 17.62 & $-14$ & 05 & 42.9 & 2314 &  6 & 3 & X     & Sa pec \\
NGC 3079 & 09 & 58 & 35.01 & $+55$ & 55 & 15.5 & 1124 & 10 & 4 & L     & SB(s)c \\
NGC 3504 & 11 & 00 & 28.53 & $+28$ & 14 & 31.3 & 1534 &  2 & 5 & SB    & SAB(s)ab \\
NGC 4151 & 12 & 08 & 01.01 & $+39$ & 41 & 02.1 &  997 &  3 & 6 & 1.5   & SAB(rs)ab\\
Mkn 231  & 12 & 54 & 05.03 & $+57$ & 08 & 38.2 & 12650 & 20& 8 & 1     & SA(rs)c?\\
Mkn 668  & 14 & 04 & 45.61 & $+28$ & 41 & 29.3 & 22957 & 40 & 9& 1     & \nodata \\
NGC 5506 & 14 & 10 & 39.14 & $-02$ & 58 & 27.1 & 1815 &  9 & 3 & 2     & Sa pec\\
\tablecomments{Properties of the sources observed in this
experiment. Coordinates mark the 21~cm continuum peak, not necessarily
the nucleus, at $\sim 2\arcsec$ resolution. References:  
(1) \protect\citeN{BSTT97}; 
(2) \protect\citeN{NW95}; 
(3) \protect\citeN{rc3}; 
(4) \protect\citeN{IS91}; 
(5) \protect\citeN{KCY93}; 
(6) \protect\citeN{PKLMABOU93}; 
(7) \protect\citeN{MWP89}; 
(8) \protect\citeN{RLWRC96};  
(9) \protect\citeN{HGdLC90}. Activity codes: $1.n=$ Seyfert $1.n$; $2=$
Seyfert 2; $2(1)= $ hidden Seyfert 1; L $=$ LINER; SB $=$ starburst; X
$=$ narrow-line X-ray galaxy (see notes in \S\protect\ref{individuals}).
\label{sourceprops}
 }
\tablenotetext{a}{This research has made use of the NASA/IPAC
  Extragalactic Database (NED) which is operated by the Jet Propulsion
  Laboratory, California Institute of Technology, under contract with
  the National Aeronautics and Space Administration.}
\enddata
\end{deluxetable}

\begin{deluxetable}{lrrrlr}
\tablecaption{Observing Parameters}
\tablewidth{0pt}
\tablehead{
\mc{Source} & \mc{$t_{int}$} & \mc{RMS} & \mc{Continuum} & \mcc{Beam} \\
 & \mc{(min)}   & \mc{(mJy beam$^{-1}$)} & \mc{(mJy beam$^{-1}$)} & \mcc{(arcsec)} 
}
\startdata
Mkn 3    &  20 & 1.5 & 895 & $2.3 \times 1.6$ & $  72\arcdeg$ \\
Mkn 6    &  52 & 1.0 & 176 & $2.1 \times 1.6$ & $ -14\arcdeg$ \\
Mkn 231  &  25 & 1.0 & 228 & $1.8 \times 1.6$ & $   9\arcdeg$ \\
Mkn 348  &  24 & 1.3 & 287 & $1.8 \times 1.5$ & $  47\arcdeg$ \\
Mkn 668  &  15 & 2.0 & 736 & $1.9 \times 1.7$ & $ -53\arcdeg$ \\
NGC 1068 & 100 & 0.9 &1395 & $2.8 \times 1.6$ & $ -23\arcdeg$ \\
NGC 2110 & 114 & 0.9 & 126 & $2.7 \times 1.5$ & $ -35\arcdeg$ \\
NGC 2639 &  60 & 0.8 &  85 & $1.8 \times 1.7$ & $  68\arcdeg$ \\
NGC 2992 &  90 & 0.7 &  49 & $3.7 \times 1.5$ & $  17\arcdeg$ \\
NGC 3079 & 141 & 0.6 &  83 & $1.7 \times 1.6$ & $  21\arcdeg$ \\
NGC 3504 &  29 & 1.4 &  52 & $1.8 \times 1.7$ & $  75\arcdeg$ \\
NGC 4151 &  49 & 0.8 & 193 & $1.9 \times 1.8$ & $  70\arcdeg$ \\
NGC 5506 &  89 & 0.7 & 223 & $2.0 \times 1.5$ & $   9\arcdeg$ \\
\label{obsparms}
\enddata
\tablecomments{The reported continuum measurement is the peak flux
density of the continuum image but not necessarily the integrated
continuum emission from the source.}
\end{deluxetable}

\begin{deluxetable}{lr@{}c@{}rr@{}c@{}rr@{}c@{}rr@{}c@{}rr@{}c@{}rr@{}c@{}r}
\tablewidth{0pt}
\tablecaption{\hi\ Absorption Properties}
\tablehead{
\mc{Source} &\mccc{Peak $\tau$} & \mccc{$\int S_{\nu}\ dv$ } & \mccc{$\int \tau_v\ dv$} & \mccc{$N_{HI}$}  & \mccc{Separation} & \mccc{Proj. Sep.}  \\
            & & & & \mccc{(Jy~\kms)}  & \mccc{(\kms)} & \mccc{($10^{21}\ {\rm cm^{-2}}$)} & \mccc{(\arcsec)} & \mccc{(pc)} \\
\mc{(1)}    & \mccc{(2)}  & \mccc{(3)} & \mccc{(4)} & \mccc{(5)} &
\mccc{(6)} & \mccc{(7)} \\
}
\startdata
Mkn 3   & $< 0.005$ & & & $< 0.09$ & & \nodata & $< 0.10$ & & \nodata & $< 0.02$ & &
  \nodata & \nodata & & \nodata & \nodata & & \nodata \\
Mkn 6\tablenotemark{b}    & 0.45 &$\pm$ &0.01 & 1.1 &
$\pm$ & 0.06 & 14.0 & $\pm$ & 1.6  &
2.6  & $\pm$ & 0.3 &  1.0   & $\pm$ &  0.10    & 
 370 & $\pm$  & 36   \\  
Mkn 231\tablenotemark{c}  & 0.073 & & 0.004 & 3.6 & & 0.17 & 33.9 &  & 3.8  &
  6.3  & & 0.7 & 0.10 & $\rightarrow$ &
  0.44 & 82 & $\rightarrow$ & 360 \\ 
Mkn 348  & $< 0.004$ & & & $< 0.08$ & & \nodata & $< 0.27$ & & \nodata & $< 0.05$ & &
  \nodata & \nodata & & \nodata & \nodata & & \nodata \\
Mkn 668  & $< 0.006$ & & & $< 0.12$ & & \nodata & $< 0.16$ & & \nodata & $< 0.03$ & &
  \nodata & \nodata & & \nodata & \nodata & & \nodata \\
NGC 1068 & 0.078 & & 0.004 & 1.5 & $\rightarrow$ & 2.9 & 2.9 & $\rightarrow$ &  15 &
  0.53 & $\rightarrow$ &  2.73 & 1.6 & $\rightarrow$ &  8.4 & 120
  &  $\rightarrow$ &  625 \\
NGC 2110 & 0.120& & 0.012& 1.0 & $\pm $ & 0.16 & 17.3 & $\pm$ & 2.1  & 2.5  &
 $\pm $ & 0.4 & 1.5 & $\pm$ & 0.16 & 230 & $\pm$ & 24 \\ 
NGC 2639 & $< 0.003$& & & $< 0.05$ & & \nodata & $< 0.56$ & & \nodata & $< 0.10$ & &
  \nodata & \nodata & & \nodata & \nodata & & \nodata \\
NGC 2992 & 0.212& &0.020 & 1.1 & & 0.12 & 31.2 & & 3.4  & 5.7  & & 0.6 & 1.0 & & 0.13 & 152 & & 19 \\
NGC 3079 & 0.788 & & 0.007 & 9.3 & & 0.19 & 135.3& & 2.3  &24.6  & & 0.5 & $<0.1$ & & \nodata &  $< 3$ & &  \nodata \\
NGC 3504 & 0.188 & & 0.026 & 1.3 & & 0.16 & 30.2 & & 3.0  & 5.5  & & 0.7 & $<0.1$ & & \nodata & $< 11$ & & \nodata \\
NGC 4151\tablenotemark{a} & 0.241 & & 0.024 & 0.9  & & 0.05  & 21.0 & & 3.0  & 3.8  & &
0.6 & $< 0.1$    & & \nodata     & $< 6$   & &\nodata    \\
NGC 5506 & 0.109& & 0.003& 3.0 & & 0.10 & 15.1 & & 0.5  & 2.6  & & 0.1 & $<0.1$ & &
  \nodata &  $<11 $ & & \nodata \\
\tablecomments{Columns: (1) Source ID; (2) peak 21~cm opacity; (3) absorption
  line strength; (4) the velocity-integrated opacity of the strongest
  absorption line; (5) the corresponding \protect\hi\ column density,
  assuming $T_S = 100$~K; (6) the angular separation between the AGN
  and the strongest \hi\ absorption; (7) the projected separation in
  parsecs. Note that the measurements of $\tau$ are necessarily lower
  limits because of confusion and unknown covering fraction. Ranges
($\rightarrow$) are given for the cases where the 
  \hi\ absorption is spatially resolved. For sources with unresolved
  \hi\ absorption or continuum, the equivalent width and column
  density are lower limits. Expressed limits are $3\sigma$, assuming a
  linewidth matching the $\sim 20$~\kms\ channel-width. For the
  NGC~4151 upper limits on \hi\ separation, we assumed
  that the AGN is located at C4W, the central VLBI
  component, after Mundell et al. (1995).
\label{hiprops}}
\tablenotetext{a}{Values include MERLIN measurements from  \citeN{MHPMKA95}. } 
\tablenotetext{b}{Values include MERLIN measurements from  \citeN{GHPM98}. } 
\tablenotetext{c}{Values include VLBA measurements from \citeN{CWU97}. } 
\enddata
\end{deluxetable}

\nocite{MHPMKA95}
\nocite{URCW98}

\begin{deluxetable}{lr@{}r@{}rr@{}r@{}rr@{}r@{}rr@{}r@{}rr@{}r@{}rr@{}r@{}r}
\tablewidth{0pt}
\tablecaption{\hi\ Absorption Line Profiles}
\tablehead{
Source & \mccc{$v_{max}$} &  \mccc{FWHM} & \mccc{$v_{50}$} &
\mccc{$v_{50}-v_{max}$} & \mccc{$v_{max}-v_{sys}$}  & \mccc{$v_{50}-v_{sys}$} \\ 
 & \mccc{(\kms)}  & \mccc{(\kms)}  & \mccc{(\kms)}  & \mccc{(\kms)}  & \mccc{(\kms)}  & \mccc{(\kms)} \\
\mc{(1)} & \mccc{(2)}  & \mccc{(3)}  & \mccc{(4)}  & \mccc{(5)}  & \mccc{(6)}  & \mccc{(7)}
}
\startdata
Mkn 6    & 5572.2 & $\pm$ & 0.6 & 57.7 & $\pm$ & 4.2 & 5575.1 & $\pm$ & 4.2 & $-$2.9  &$\pm$ & 2.6 & $-$67.8 &$\pm$ & 10.0 &$-$64.9 & $\pm$& 4.9\\
Mkn 231  &12664.2 & & 5.9 &211.8 & & 6.8 &12647.5 & & 6.8 & 16.7 & &
6.5 &  14.2 & & 20.9 & $-$2.5 & & 9.4 \\
NGC 1068 & 1117.8 & & 1.5 &161.1 &  & 5.0 & 1170.2 &  & 5.0 & 52.4  &  & 4.7 & $-$30.2 & &  5.2 &22.2  &  & 6.8\\
NGC 2110 & 2626.3 & & 4.3 & 88.5 & & 7.6 & 2616.3 & & 7.6 & 10.0 & & 5.6 & 291.3 & & 20.5 &281.3 & & 9.4\\
NGC 2992 & 2296.5 & & 0.9 & 95.0 & & 4.8 & 2326.6 & & 4.8 & $-$30.1  & & 4.5 & $-$17.5 & &  6.1 & 12.6 & & 6.5\\
NGC 3079 & 1136.2 & & 0.4 &132.5 & & 1.4 & 1130.0 & & 1.4 & 6.2  & & 1.2 &  12.2 & & 10.0 &  6.0 & & 1.8\\
NGC 3504 & 1542.0 & & 1.0 &135.5 & & 7.4 & 1525.3 & & 7.4 & 16.7 & & 6.0 &   8.0 & &  2.2 & $-$8.7 & & 9.5\\
NGC 4151 &  994.7 & & 0.9 & 89.8 & & 5.2 &  991.0 & & 5.2 & 3.7  & & 3.3 &  $-$2.3 & &  3.1 & $-$6.0 & & 6.1\\
NGC 5506 & 1802.3 & & 0.8 & 95.0 & & 3.5 & 1821.7 & & 3.5 & $-$19.4  & & 3.5 & $-$12.7 & &  9.0 &  6.7 & & 5.0\\
\tablecomments{Columns: (1) Source
  ID; (2) velocity at absorption line minimum (i.e., peak line
  strength); (3) full-width at half maximum; (4) the velocity mean
  across the FWHM; (5) as a line asymmetry parameter, the difference
  between $v_{max}$ and $v_{50}$, with positive values indicating a
  red asymmetry in the line wings; (6) \& (7) the difference between
  the systemic velocity and the line profile velocities, with positive
  values indicating a red-shift relative to systemic.\label{hikin}}
\enddata
\end{deluxetable}

\begin{deluxetable}{lr@{.}l@{}lc}
\tablewidth{0pt}
\tablecaption{Soft X-ray Absorbing Columns}
\tablehead{
Source & \mccc{$N_H$} & Reference \\
       & \mccc{($ 10^{21}$ cm\mtwo)} & 
}
\startdata
Mrk 3    & 707 & 9 & $^{+51}_{-47}$ &  (1) \\
Mrk 6    &  38 & 0 & $^{+5.0}_{-0.6}$ & (2) \\
Mrk 231  &   5 & 3 & $^{+3.3}_{-3.3}$ & (3) \\
Mrk 348  & 112 & 2 & $^{+20}_{-21}$ & (1) \\
Mrk 501  & $<18$ &  &  & (4) \\
NGC 1068 & $>1000$ &  &  & (5) \\
NGC 2110 &  28 & 4 & $^{+1.9}_{-1.6}$ & (6) \\
NGC 2992 &   1 & 3 & $^{+1.6}_{-0.7}$ & (6)\\
NGC 3079 &   0 & 6 & $^{+0.1}_{-0.1}$ & (7)\\
NGC 3227 &   5 & 0 & $^{+0.9}_{-0.9}$ & (8) \\
NGC 4151 &  22 & 6 & $^{+1.0}_{-1.0}$ & (8) \\
NGC 5506 &  21 & 9 & $^{+2.6}_{-2.4}$ & (9)  \\
\tablecomments{For
  those cases where more than one X-ray spectral model is available in the
  literature, we selected the model giving a reasonable fit to the
  data and the best constraint on the hydrogen column. References: (1)
  Mulchaey, Mushotzky, \& Weaver (1992); (2) Feldmeier et al. (1999);
  (3) Wang et al. (1996); (4) Mushotzky et al. (1978); (5) Matt et
  al. (1997) and references therein; (6) Turner
  et al. (1997); (7) Read, Ponman, \& Strickland (1997); (8) George et
  al. (1998); (9) Mulchaey et al. (1993). 
  \label{tab-cx}} 
\enddata
\end{deluxetable}

\nocite{Metal97}
\nocite{MMW92}
\nocite{FBEFIM99}
\nocite{WBB96}
\nocite{MBHSSRP78}
\nocite{TGNM97}
\nocite{GTNNMY98}
\nocite{RPS97}
\nocite{MCWMW93}

\begin{deluxetable}{lrrr@{}rr@{}rr@{}rr@{}rrr}
\tablewidth{0pt}
\tablecaption{Model Disk Parameters}
\tablehead{
\mc{Source} &
\mc{$i$} &
\mc{PA} &
\mcc{$v_{rot}$} & 
\mcc{$N_{HI}$(face-on)} & 
\mcc{$r_{in}$} & 
\mcc{$r_{out}$} & 
\mc{$\log{{\rm M}_{HI}}$} &
\mc{$\chi^2$} \\ 
 &
\mc{(deg)} &
\mc{(deg)} &
\mcc{(\kms)} & 
\mcc{($\times 10^{21}$~cm\mtwo)} & 
\mcc{(pc)} & 
\mcc{(pc)} & 
\mc{(\Msun)} &
   \\ 
\mc{(1)} &
\mc{(2)} &
\mc{(3)} &
\mcc{(4)} & 
\mcc{(5)} &
\mcc{(6)} & 
\mcc{(7)} & 
\mc{(8)} &
\mc{(9)}}
\startdata
Mkn 6     & 51  & 130 & 140 & $\pm 4$ & 1.65 &$\pm 0.07$ & 293 & \nodata & 426 & \nodata & 6.60 & 143.8 \\
NGC 1068  & 40  & $-$45  & 320  & 82  &  0.82 & 0.056 & 212&  \nodata & 957&  \nodata & 7.26 &  64.5\\ 
NGC 2110  & 53  & 161  & 412  &  5  &  2.31 & 0.25 & 250&  \nodata & 552&  \nodata & 7.15 &  57.0\\
NGC 2992  & 70  &$-$154  & 147  &  8  &  0.73 & 0.05 &  36& $\pm$34 & 921&  \nodata & 7.20 & 185.9\\
NGC 3079  & 84  & 166  & 177  &  1  & 1.53 & 0.01 &   8&  1 & 282&  $\pm$4 & 6.48 & 1089.0\\
NGC 3504  & 22  & $-$33  & 235  & 27  &  2.33 & 0.21 &   8&  1 & $<312$ & \nodata & $< 6.76$ &  50.5\\
NGC 4151\tablenotemark{a}                                                                 
          & 21  &  22 & $-$148  &  8  &  9.64 & 0.63 &   \nodata &
          \nodata &  7 &  \nodata &  4.85 & 157.8\\
NGC 4151\tablenotemark{b}                                                                 
          & 50  & 180  & 839  &122  & 10.20 & 0.69 &   \nodata &
          \nodata &  10 &  \nodata &  4.87 &  66.6\\
NGC 5506  & 80  & $-$87  & 181  &  5   &  0.25 & 0.01 &   7&  2 & 406&
26 &  6.00 & 963.2 \\
\tablecomments{Columns are: (1) Source ID; (2)
  inclination; (3) position angle to receding major axis; (4) the
  deprojected rotational velocity at 100~pc radius; (5) face$-$on
  column density through the disk; (6) the deprojected inner and (7)
  outer radii; (8) logarithm of the \hi\ mass (in \Msun) assuming a
  uniform and symmetric disk; and (9) $\chi^2$. Error-bars are given
  for fitted parameters. Where not explicitly fit, the inner and outer
  radii span the deprojected range of the observed absorption.  The
  reported $\chi^2$ values are for 50 spectral data points and 3--5
  fitted parameters. We fit the MERLIN data of Gallimore et al. (1998)
  for Mkn~6; the $\chi^2$ values are for 53 spectral data points and
  three fitted parameters.  Note that the \protect\hi\ masses of the
  model disks are less than 0.2\% of the dynamical mass at the
  corresponding radii. References for the kinematic axes are, Mkn~6:
  Meaburn, Whitehead, \& Pedlar (1989); NGC~1068: Brinks et al.
  (1997) and Gallimore et al.  (1994); NGC~2110: Wilson \& Baldwin
  (1985); NGC~2992: Colina et al.  (1987); NGC~3079: Irwin \& Seaquist
  (1991); NGC~3504: Kenney, Carlstrom, \& Young (1993); NGC~4151:
  Pedlar et al. (1992) (galaxy disk) and Mundell et al. (1995) (AGN
  disk); NGC~5506: Maiolino et al. (1994).
\label{t-diskfits}}
\tablenotetext{a}{The inclination and position angle are fixed to
  those of the outer galaxy disk.}
\tablenotetext{b}{The inclination and position angle are fixed to
  those predicted for an AGN disk (i.e., perpendicular to the radio jet).}
\enddata
\end{deluxetable}

\nocite{GBOBP94}
\nocite{BSTT97}
\nocite{WB85}
\nocite{CFKP87}
\nocite{IS91}
\nocite{KCY93}
\nocite{PHAU92}
\nocite{MHPMKA95}
\nocite{MSSR94}
\nocite{MWP89}

\begin{deluxetable}{lr@{}rr@{}rrr}
\tablewidth{0pt}
\tablecaption{Streaming Velocity Models}
\tablehead{
\mc{Source} & \mcc{$v_{rad}$} & \mcc{$\dot{\rm M}$} & \mc{$\chi^2$} & \mc{$\Delta\chi^2$} \\
            & \mcc{(\kms)}    & \mcc{ (\Msun yr\mone)} & & \\
\mc{(1)} & \mcc{(2)} & \mcc{(3)} & \mc{(4)}& \mc{(5)} }
\startdata
Mkn 6    &  $<59.0$ & \nodata & $< 0.57$ & \nodata & 95.4 &  48.4\\
NGC 1068 &  $-$46.6 & $\pm$11.2 & $-0.90$ & $\pm 0.21$ &    36.8 & 27.7 \\
NGC 2110 &   47.3 & 11.7 & 1.16 & 0.22 &   42.4      & 14.6 \\
NGC 2992 &   37.6 &  5.4 & 0.67 & 0.09 &  131.8      & 54.1 \\
NGC 3079 &   16.6 &  0.6 & 0.18 & 0.01 &  927.1      & 161.9 \\
NGC 3504 &    9.6 & 19.4 & 0.18 & 0.36 &   50.5      & 0.3 \\
NGC 4151\tablenotemark{a} & $-0.32$ & 0.04 & $-$182.4 & 22.5 &   48.5 & 109.3 \\
NGC 4151\tablenotemark{b} & $-0.07$ & 0.01 &  $-$24.8 &  3.8 &   46.1 & 20.5 \\
NGC 5506 &   19.3 &  0.9  & 0.05 & 0.01 & 715.8 & 247.4  \\
\tablecomments{Streaming velocities based on disk model fitting to the
  \hi\ absorption profiles and locations. Positive velocities imply
  motion toward the center of the galaxy. All models also include a
  rotational component, but the fitted values are not significantly
  different from those listed in Table~\protect\ref{t-diskfits}. The
  exception is Mkn~6, for which the fitted rotational velocity drops
  to $\sim 40$~\kms; the fitted value therefore places a limit
  appropriate for gas dominated by radial motion. The columns are (1)
  Source ID; (2) streaming velocity; (3) corresponding mass infall
  rate; (4) $\chi^2$; and (5) $\Delta\chi^2$, the improvement in
  $\chi^2$ relative to the purely rotational model
  (Table~\protect\ref{t-diskfits}). \label{t-stream}}
\tablenotetext{a}{The inclination and position angle are fixed to
  those of the outer galaxy disk.} \tablenotetext{b}{The inclination and
  position angle are fixed to those predicted for an AGN disk (i.e.,
  perpendicular to the radio jet).}
\enddata
\end{deluxetable}


\begin{thebibliography}{}

\bibitem[\protect\citeauthoryear{{Antonucci}}{{Antonucci}}{1993}]{Antonucci93}
{Antonucci}, R. 1993, ARAA, 31, 473

\bibitem[\protect\citeauthoryear{{Antonucci} \& {Miller}}{{Antonucci} \&
  {Miller}}{1985}]{AM85}
{Antonucci}, R. R.~J.,  \& {Miller}, J.~S. 1985, ApJ, 297, 621

\bibitem[\protect\citeauthoryear{{Athanassoula}}{{Athanassoula}}{1992}]{Athana%
ssoula92}
{Athanassoula}, E. 1992, MNRAS, 259, 345

\bibitem[\protect\citeauthoryear{{Athanassoula}}{{Athanassoula}}{1994}]{Athana%
ssoula94p143}
{Athanassoula}, E. 1994, in Mass Transfer Induced Activity in Galaxies, ed.
  I.~{Shlosman} (Cambridge: Cambridge University Press), 143

\bibitem[\protect\citeauthoryear{{Awaki} et~al.}{{Awaki} et~al.}{1991}]{AKIH91}
{Awaki}, H., {Koyama}, K., {Inoue}, H.,  \& {Halpern}, J.~P. 1991, PASJ, 43,
  195

\bibitem[\protect\citeauthoryear{{Baan} \& {Haschick}}{{Baan} \&
  {Haschick}}{1983}]{BH83}
{Baan}, W.~A.,  \& {Haschick}, A.~D. 1983, AJ, 88, 1088

\bibitem[\protect\citeauthoryear{{Baan} \& {Irwin}}{{Baan} \&
  {Irwin}}{1995}]{BI95}
{Baan}, W.~A.,  \& {Irwin}, J.~A. 1995, ApJ, 446, 602

\bibitem[\protect\citeauthoryear{{Baars} et~al.}{{Baars} et~al.}{1977}]{BGPW77}
{Baars}, J. W.~M., {Genzel}, R., {Pauliny-Toth}, I. I.~K.,  \& {Witzel}, A.
  1977, A\&A, 619, 99

\bibitem[\protect\citeauthoryear{{Bahcall} \& {Ekers}}{{Bahcall} \&
  {Ekers}}{1969}]{BE69}
{Bahcall}, J.~N.,  \& {Ekers}, R.~D. 1969, ApJ, 157, 1055

\bibitem[\protect\citeauthoryear{{Begeman}}{{Begeman}}{1987}]{Begeman87}
{Begeman}, K.~G. 1987, Ph. D. Thesis (Groningen, the Netherlands: University of
  Groningen)

\bibitem[\protect\citeauthoryear{{Bosma}}{{Bosma}}{1978}]{Bosma78}
{Bosma}, A. 1978, Ph. D. Thesis (Groningen, the Netherlands: University of
  Groningen)

\bibitem[\protect\citeauthoryear{{Braun}}{{Braun}}{1997}]{Braun97}
{Braun}, R. 1997, ApJ, 484, 637

\bibitem[\protect\citeauthoryear{{Brinks} et~al.}{{Brinks}
  et~al.}{1997}]{BSTT97}
{Brinks}, E., {Skillman}, E.~D., {Terlevich}, R.~J.,  \& {Terlevich}, E. 1997,
  Ap\&SS, 248, 23

\bibitem[\protect\citeauthoryear{{Broeils} \& {van Woerden}}{{Broeils} \& {van
  Woerden}}{1994}]{BvW94}
{Broeils}, A.~H.,  \& {van Woerden}, H. 1994, A\&AS, 107, 129

\bibitem[\protect\citeauthoryear{{Bryant} \& {Scoville}}{{Bryant} \&
  {Scoville}}{1996}]{BS96}
{Bryant}, P.~M.,  \& {Scoville}, N.~Z. 1996, ApJ, 457, 678

\bibitem[\protect\citeauthoryear{{Carilli}}{{Carilli}}{1991}]{Carilli91}
{Carilli}, C.~L. 1991, VLA Test Memorandum No. 158 (Socorro: NRAO)

\bibitem[\protect\citeauthoryear{{Carilli}, {Wrobel}, \& {Ulvestad}}{{Carilli}
  et~al.}{1997}]{CWU97}
{Carilli}, C.~L., {Wrobel}, J.~M.,  \& {Ulvestad}, J.~S. 1997, AJ, 115, 928

\bibitem[\protect\citeauthoryear{{Carral}, {Turner}, \& {Ho}}{{Carral}
  et~al.}{1990}]{CTH90}
{Carral}, P., {Turner}, J.~L.,  \& {Ho}, P. T.~P. 1990, 362, 290, 434

\bibitem[\protect\citeauthoryear{{Clements}}{{Clements}}{1981}]{Clements81}
{Clements}, E.~D. 1981, MNRAS, 197, 829

\bibitem[\protect\citeauthoryear{{Colbert} et~al.}{{Colbert}
  et~al.}{1996}]{CBGOC96}
{Colbert}, E. J.~M., {Baum}, S.~A., {Gallimore}, J.~F., {O'Dea}, C.~P.,  \&
  {Christensen}, J.~A. 1996, ApJ, 467, 551

\bibitem[\protect\citeauthoryear{{Colbert,} et~al.}{{Colbert,}
  et~al.}{1998}]{CBOV98}
{Colbert,}, E. J.~M., {Baum}, S.~A., {O'Dea}, C.~P.,  \& {Veilleux}, S. 1998,
  ApJ, 496, 786

\bibitem[\protect\citeauthoryear{{Cole} et~al.}{{Cole} et~al.}{1998}]{CPMGH98}
{Cole}, G. H.~J., {Pedlar}, A., {Mundell}, C.~G., {Gallimore}, J.~F.,  \&
  {Holloway}, A.~J. 1998, MNRAS, 301, 782

\bibitem[\protect\citeauthoryear{{Colina} et~al.}{{Colina}
  et~al.}{1987}]{CFKP87}
{Colina}, L., {Fricke}, K.~J., {Kollatschny}, W.,  \& {Perryman}, M. A.~C.
  1987, A\&A, 178, 51

\bibitem[\protect\citeauthoryear{{Condon} \& {Broderick}}{{Condon} \&
  {Broderick}}{1988}]{CB88}
{Condon}, J.~J.,  \& {Broderick}, J.~J. 1988, AJ, 96, 30

\bibitem[\protect\citeauthoryear{{Cowie}}{{Cowie}}{1981}]{Cowie81}
{Cowie}, L.~L. 1981, ApJ, 245, 66

\bibitem[\protect\citeauthoryear{{de Vaucouleurs} et~al.}{{de Vaucouleurs}
  et~al.}{1991}]{rc3}
{de Vaucouleurs}, G., {de Vaucouleurs}, A., {Corwin}, H.~G., {Buta}, R.~J.,
  {Paturel}, G.,  \& {Fouqu\'e}, P. 1991, Third Reference Catalogue of Bright
  Galaxies (Berlin: Springer-Verlag)

\bibitem[\protect\citeauthoryear{{Dickey}}{{Dickey}}{1986}]{Dickey86}
{Dickey}, J.~M. 1986, ApJ, 300, 190

\bibitem[\protect\citeauthoryear{{Dickey}, {Terzian}, \& {Salpeter}}{{Dickey}
  et~al.}{1978}]{DTS78}
{Dickey}, J.~M., {Terzian}, Y.,  \& {Salpeter}, E.~E. 1978, ApJS, 36, 77

\bibitem[\protect\citeauthoryear{{Eckart} \& {Genzel}}{{Eckart} \&
  {Genzel}}{1996}]{EG96}
{Eckart}, A.,  \& {Genzel}, R. 1996, Nature, 383, 415

\bibitem[\protect\citeauthoryear{{Elmegreen} et~al.}{{Elmegreen}
  et~al.}{1997}]{ECSM97}
{Elmegreen}, D.~M., {Chromey}, F.~R., {Santos}, M.,  \& {Marshall}, D. 1997,
  AJ, 114, 1850

\bibitem[\protect\citeauthoryear{{Elvis} et~al.}{{Elvis} et~al.}{1990}]{EFWB90}
{Elvis}, M., {Fassnacht}, C., {Wilson}, A.~S.,  \& {Briel}, U. 1990, ApJ, 361,
  459

\bibitem[\protect\citeauthoryear{{Feldmeier} et~al.}{{Feldmeier}
  et~al.}{1999}]{FBEFIM99}
{Feldmeier}, J., {Brandt}, W.~N., {Elvis}, M., {Fabian}, A.~C., {Iwasawa}, K.,
  \& {Mathur}, S. 1999, ApJ, 510, 167

\bibitem[\protect\citeauthoryear{{Ferland} \& {Osterbrock}}{{Ferland} \&
  {Osterbrock}}{1986}]{FO86}
{Ferland}, G.~J.,  \& {Osterbrock}, D.~E. 1986, ApJ, 300, 658

\bibitem[\protect\citeauthoryear{{Field}}{{Field}}{1959a}]{Field59a}
{Field}, G.~B. 1959a, ApJ, 129, 525

\bibitem[\protect\citeauthoryear{{Field}}{{Field}}{1959b}]{Field59b}
{Field}, G.~B. 1959b, ApJ, 129, 536

\bibitem[\protect\citeauthoryear{{Field}}{{Field}}{1959c}]{Field59c}
{Field}, G.~B. 1959c, ApJ, 129, 551

\bibitem[\protect\citeauthoryear{{Friedli} \& {Benz}}{{Friedli} \&
  {Benz}}{1993}]{FB93}
{Friedli}, D.,  \& {Benz}, W. 1993, A\&A, 268, 65

\bibitem[\protect\citeauthoryear{{Gallimore}, {Baum}, \& {O'Dea}}{{Gallimore}
  et~al.}{1996}]{n1068b}
{Gallimore}, J.~F., {Baum}, S.~A.,  \& {O'Dea}, C.~P. 1996, ApJ, 464, 198

\bibitem[\protect\citeauthoryear{{Gallimore}, {Baum}, \& {O'Dea}}{{Gallimore}
  et~al.}{1997}]{GBO97}
{Gallimore}, J.~F., {Baum}, S.~A.,  \& {O'Dea}, C.~P. 1997, Nature, 388, 852

\bibitem[\protect\citeauthoryear{{Gallimore} et~al.}{{Gallimore}
  et~al.}{1994}]{GBOBP94}
{Gallimore}, J.~F., {Baum}, S.~A., {O'Dea}, C.~P., {Brinks}, E.,  \& {Pedlar},
  A. 1994, ApJ, 422, L13

\bibitem[\protect\citeauthoryear{{Gallimore} et~al.}{{Gallimore}
  et~al.}{1996a}]{maserpaper}
{Gallimore}, J.~F., {Baum}, S.~A., {O'Dea}, C.~P., {Brinks}, E.,  \& {Pedlar},
  A. 1996a, ApJ, 462, 740

\bibitem[\protect\citeauthoryear{{Gallimore} et~al.}{{Gallimore}
  et~al.}{1996b}]{n1068a}
{Gallimore}, J.~F., {Baum}, S.~A., {O'Dea}, C.~P.,  \& {Pedlar}, A. 1996b, ApJ,
  458, 136

\bibitem[\protect\citeauthoryear{{Gallimore} et~al.}{{Gallimore}
  et~al.}{1998}]{GHPM98}
{Gallimore}, J.~F., {Holloway}, A.~J., {Pedlar}, A.,  \& {Mundell}, C.~G. 1998,
  A\&A, 333, 13

\bibitem[\protect\citeauthoryear{{Genzel} et~al.}{{Genzel}
  et~al.}{1985}]{GWCT85}
{Genzel}, R., {Watson}, D.~M., {Crawford}, M.~K.,  \& {Townes}, C.~H. 1985,
  ApJ, 297, 766

\bibitem[\protect\citeauthoryear{{George} et~al.}{{George}
  et~al.}{1998}]{GTNNMY98}
{George}, I.~M., {Turner}, T.~J., {Netzer}, H., {Nandra}, K., {Mushotzky},
  R.~F.,  \& {Yaqoob}, T. 1998, ApJS, 114, 73

\bibitem[\protect\citeauthoryear{{Giovanelli} \& {Haynes}}{{Giovanelli} \&
  {Haynes}}{1988}]{GH88p522}
{Giovanelli}, R.,  \& {Haynes}, M.~P. 1988, in Galactic and Extragalactic Radio
  Astronomy, ed. G.~L. {Verschuur} \& K.~I. {Kellermann} (Berlin:
  Springer-Verlag), 522

\bibitem[\protect\citeauthoryear{{Halkides}, {Ulvestad}, \& {Roy}}{{Halkides}
  et~al.}{1997}]{HUR97}
{Halkides}, D., {Ulvestad}, J.,  \& {Roy}, A. 1997, BAAS, 191, 104.04

\bibitem[\protect\citeauthoryear{{Harrison} et~al.}{{Harrison}
  et~al.}{1986}]{HPUBGP86}
{Harrison}, B., {Pedlar}, A., {Unger}, S.~W., {Burgess}, P., {Graham}, D.~A.,
  \& {Preuss}, E. 1986, MNRAS, 218, 775

\bibitem[\protect\citeauthoryear{{Heckman}}{{Heckman}}{1980}]{Heckman80}
{Heckman}, T.~M. 1980, A\&A, 87, 152

\bibitem[\protect\citeauthoryear{{Heckman}, {Balick}, \& {Sullivan}}{{Heckman}
  et~al.}{1978}]{HBS78}
{Heckman}, T.~M., {Balick}, B.,  \& {Sullivan}, W. T.~{\sc III}. 1978, ApJ,
  224, 745

\bibitem[\protect\citeauthoryear{{Heckman} et~al.}{{Heckman}
  et~al.}{1982}]{HSBS82}
{Heckman}, T.~M., {Sancisi}, R., {Balick}, B.,  \& {Sullivan}, W. T.~{\sc III}.
  1982, MNRAS, 199, 425

\bibitem[\protect\citeauthoryear{{Helfer} \& {Blitz}}{{Helfer} \&
  {Blitz}}{1995}]{HB95}
{Helfer}, T.~T.,  \& {Blitz}, L. 1995, ApJ, 450, 90

\bibitem[\protect\citeauthoryear{{Helfer} \& {Blitz}}{{Helfer} \&
  {Blitz}}{1997}]{HB97}
{Helfer}, T.~T.,  \& {Blitz}, L. 1997, ApJ, 478, 162

\bibitem[\protect\citeauthoryear{{Helfer} et~al.}{{Helfer}
  et~al.}{1998}]{HTRSVWBB98}
{Helfer}, T.~T., {Thornley,}, M.~D., {Regan}, M.~W., {Sheth}, K., {Vogel},
  S.~N., {Wong}, T., {Blitz}, L.,  \& {Bock}, D. 1998, BAAS, 192, 7301

\bibitem[\protect\citeauthoryear{{Heller} \& {Shlosman}}{{Heller} \&
  {Shlosman}}{1994}]{HS94}
{Heller}, C.~H.,  \& {Shlosman}, I. 1994, ApJ, 424, 84

\bibitem[\protect\citeauthoryear{{Herrnstein} et~al.}{{Herrnstein}
  et~al.}{1998}]{HGMDINM98}
{Herrnstein}, J.~R., {Greenhill}, L.~J., {Moran}, J.~M., {Diamond}, P.~J.,
  {Inoue}, M., {Nakai}, N.,  \& {Miyoshi}, M. 1998, ApJ, 497, 89

\bibitem[\protect\citeauthoryear{{Ho}, {Filippenko}, \& {Sargent}}{{Ho}
  et~al.}{1997}]{HFS97}
{Ho}, L.~C., {Filippenko}, A.~V.,  \& {Sargent}, W. L.~W. 1997, ApJ, 487, 591

\bibitem[\protect\citeauthoryear{{Huchra} et~al.}{{Huchra}
  et~al.}{1990}]{HGdLC90}
{Huchra}, J., {Gellar}, M.~J., {de Lapparent}, V.,  \& {Corwin}, H. G.~J. 1990,
  ApJS, 72, 433

\bibitem[\protect\citeauthoryear{{Irwin} \& {Seaquist}}{{Irwin} \&
  {Seaquist}}{1990}]{IS90}
{Irwin}, J.~A.,  \& {Seaquist}, E.~R. 1990, ApJ, 353, 469

\bibitem[\protect\citeauthoryear{{Irwin} \& {Seaquist}}{{Irwin} \&
  {Seaquist}}{1991}]{IS91}
{Irwin}, J.~A.,  \& {Seaquist}, E.~R. 1991, ApJ, 371, 111

\bibitem[\protect\citeauthoryear{{Jackson} et~al.}{{Jackson}
  et~al.}{1996}]{JHPB96}
{Jackson}, J.~M., {Heyer}, M.~H., {Paglione}, T. A.~D.,  \& {Bolatto}, A.~D.
  1996, ApJ, 456, 91

\bibitem[\protect\citeauthoryear{{Keel}}{{Keel}}{1980}]{Keel80}
{Keel}, W.~C. 1980, AJ, 85, 198

\bibitem[\protect\citeauthoryear{{Kenney}, {Carlstrom}, \& {Young}}{{Kenney}
  et~al.}{1993}]{KCY93}
{Kenney}, J. D.~P., {Carlstrom}, J.~E.,  \& {Young}, J.~S. 1993, ApJ, 418, 687

\bibitem[\protect\citeauthoryear{{Kennicutt}}{{Kennicutt}}{1989}]{Kennicutt89}
{Kennicutt}, R.~C. 1989, ApJ, 344, 685

\bibitem[\protect\citeauthoryear{{Kennicutt}}{{Kennicutt}}{1990}]{Kennicutt90}
{Kennicutt}, R.~C. 1990, ApJ, 498, 541

\bibitem[\protect\citeauthoryear{{Kormendy} \& {Richstone}}{{Kormendy} \&
  {Richstone}}{1995}]{KR95}
{Kormendy}, J.,  \& {Richstone}, D. 1995, ARA\&A, 33, 581

\bibitem[\protect\citeauthoryear{{Koski}}{{Koski}}{1978}]{Koski78}
{Koski}, A. 1978, ApJ, 223, 56

\bibitem[\protect\citeauthoryear{{Kriss} et~al.}{{Kriss}
  et~al.}{1992}]{Ketal92}
{Kriss}, G.~A.,  et~al. 1992, ApJ, 392, 485

\bibitem[\protect\citeauthoryear{{Krolik} \& {Lepp}}{{Krolik} \&
  {Lepp}}{1989}]{KL89}
{Krolik}, J.~H.,  \& {Lepp}, S. 1989, ApJ, 347, 179

\bibitem[\protect\citeauthoryear{{Kukula} et~al.}{{Kukula}
  et~al.}{1999}]{KGPS99}
{Kukula}, M.~J., {Ghosh}, T., {Pedlar}, A.,  \& {Schilizzi}, R.~T. 1999, ApJ,
  in press

\bibitem[\protect\citeauthoryear{{Kukula} et~al.}{{Kukula}
  et~al.}{1993}]{Kukula93}
{Kukula}, M.~J., {Ghosh}, T., {Pedlar}, A., {Schillizzi}, R.~T., {Miley},
  G.~K., {de~Bruyn}, A.~G.,  \& {Saikia}, D.~J. 1993, MNRAS, 264, 893

\bibitem[\protect\citeauthoryear{{Kukula} et~al.}{{Kukula}
  et~al.}{1996}]{Kukula96}
{Kukula}, M.~J., {Holloway}, A.~J., {Pedlar}, A., {Meaburn}, J., {Lopez},
  J.~A., {Axon}, D.~J., {Schilizzi}, R.~T.,  \& {Baum}, S.~A. 1996, MNRAS, 280,
  1283

\bibitem[\protect\citeauthoryear{{Lawrence}}{{Lawrence}}{1987}]{Lawrence87}
{Lawrence}, A. 1987, PASP, 99, 309

\bibitem[\protect\citeauthoryear{{Lawrence}}{{Lawrence}}{1991}]{Lawrence91}
{Lawrence}, A. 1991, MNRAS, 252, 586

\bibitem[\protect\citeauthoryear{{Lilie}}{{Lilie}}{1994}]{Lilie94}
{Lilie}, P. 1994, VLA Test Memorandum No. 190 (Socorro: NRAO)

\bibitem[\protect\citeauthoryear{{Liszt} et~al.}{{Liszt}
  et~al.}{1983}]{LvdHBO83}
{Liszt}, H.~S., {van der Hulst}, J.~M., {Burton}, W.~B.,  \& {Ondrechen}, M.~P.
  1983, A\&A, 126, 341

\bibitem[\protect\citeauthoryear{{Maiolino} et~al.}{{Maiolino}
  et~al.}{1994}]{MSSR94}
{Maiolino}, R., {Stanga}, R., {Salvati}, M.,  \& {Rodriguez~Espinosa}, J.~M.
  1994, A\&A, 290, 40

\bibitem[\protect\citeauthoryear{{Malkan}, {Gorjian}, \& {Tam}}{{Malkan}
  et~al.}{1998}]{MGT98}
{Malkan}, M.~A., {Gorjian}, V.,  \& {Tam}, R. 1998, ApJS, 117, 25

\bibitem[\protect\citeauthoryear{{Maloney}, {Hollenbach}, \&
  {Tielens}}{{Maloney} et~al.}{1996}]{MHT96}
{Maloney}, P.~R., {Hollenbach}, D.~J.,  \& {Tielens}, A. G. G.~M. 1996, ApJ,
  466, 561

\bibitem[\protect\citeauthoryear{{Martin}}{{Martin}}{1998}]{Martin98}
{Martin}, M.~C. 1998, A\&AS, 131, 73

\bibitem[\protect\citeauthoryear{{Matt} et~al.}{{Matt} et~al.}{1997}]{Metal97}
{Matt}, G.,  et~al. 1997, A\&A, 325, L13

\bibitem[\protect\citeauthoryear{{McClintock} et~al.}{{McClintock}
  et~al.}{1979}]{MRCVvP79}
{McClintock}, J.~E., {Remillard}, R.~A., {Canizares}, C.~R., {Veron}, P.,  \&
  {van Paradijs}, J. 1979, ApJ, 233, 809

\bibitem[\protect\citeauthoryear{{Meaburn}, {Whitehead}, \& {Pedlar}}{{Meaburn}
  et~al.}{1989}]{MWP89}
{Meaburn}, J., {Whitehead}, M.~J.,  \& {Pedlar}, A. 1989, MNRAS, 241, 1P

\bibitem[\protect\citeauthoryear{{Mezger} \& {Henderson}}{{Mezger} \&
  {Henderson}}{1967}]{MH67}
{Mezger}, P.~G.,  \& {Henderson}, A.~P. 1967, ApJ, 147, 471

\bibitem[\protect\citeauthoryear{{Miller} \& {Goodrich}}{{Miller} \&
  {Goodrich}}{1990}]{MG90}
{Miller}, J.~S.,  \& {Goodrich}, R.~W. 1990, ApJ, 355, 456

\bibitem[\protect\citeauthoryear{{Miyoshi} et~al.}{{Miyoshi}
  et~al.}{1995}]{Miyoshi95}
{Miyoshi}, M., {Moran}, J., {Herrnstein}, J., {Greenhill}, L., {Nakai}, N.,
  {Diamond}, P.,  \& {Inoue}, M. 1995, Nature, 373, 127

\bibitem[\protect\citeauthoryear{{Moran} et~al.}{{Moran} et~al.}{1992}]{MHBB92}
{Moran}, E.~C., {Halpern}, J.~C., {Bothun}, G.~D.,  \& {Becker}, R.~H. 1992,
  AJ, 104, 990

\bibitem[\protect\citeauthoryear{{Morrison} \& {McCammon}}{{Morrison} \&
  {McCammon}}{1983}]{MM83}
{Morrison}, R.,  \& {McCammon}, D. 1983, ApJ, 270, 199

\bibitem[\protect\citeauthoryear{{Mulchaey} et~al.}{{Mulchaey}
  et~al.}{1993}]{MCWMW93}
{Mulchaey}, J.~S., {Colbert}, E., {Wilson}, A.~S., {Mushotzky}, R.~F.,  \&
  {Weaver}, K.~A. 1993, ApJ, 414, 144

\bibitem[\protect\citeauthoryear{{Mulchaey}, {Mushotzky}, \&
  {Weaver}}{{Mulchaey} et~al.}{1992}]{MMW92}
{Mulchaey}, J.~S., {Mushotzky}, R.~F.,  \& {Weaver}, K.~A. 1992, ApJ, 390, L69

\bibitem[\protect\citeauthoryear{{Mulchaey} \& {Regan}}{{Mulchaey} \&
  {Regan}}{1997}]{MR97}
{Mulchaey}, J.~S.,  \& {Regan}, M.~W. 1997, ApJ, 482, L135

\bibitem[\protect\citeauthoryear{{Mundell} et~al.}{{Mundell}
  et~al.}{1995}]{MHPMKA95}
{Mundell}, C.~G., {Holloway}, A.~J., {Pedlar}, A., {Meaburn}, J., {Kukula},
  M.~J.,  \& {Axon}, D.~J. 1995, MNRAS, 275, 67

\bibitem[\protect\citeauthoryear{{Mushotzky} et~al.}{{Mushotzky}
  et~al.}{1978}]{MBHSSRP78}
{Mushotzky}, R.~F., {Boldt}, E.~A., {Holt}, S.~S., {Serlemitsos}, P.~J.,
  {Swank}, J.~H., {Rothschild}, R.~H.,  \& {Pravdo}, S.~H. 1978, ApJ, 226, L65

\bibitem[\protect\citeauthoryear{{Nagar} et~al.}{{Nagar} et~al.}{1999}]{NWMG99}
{Nagar}, N.~M., {Wilson}, A.~S., {Mulchaey}, J.~S.,  \& {Gallimore}, J.~F.
  1999, ApJS, in press

\bibitem[\protect\citeauthoryear{{Neff} \& {de~Bruyn}}{{Neff} \&
  {de~Bruyn}}{1983}]{NdB83}
{Neff}, S.~G.,  \& {de~Bruyn}, A.~G. 1983, A\&A, 128, 318

\bibitem[\protect\citeauthoryear{{Nelson} \& {Whittle}}{{Nelson} \&
  {Whittle}}{1995}]{NW95}
{Nelson}, C.~H.,  \& {Whittle}, M. 1995, ApJS, 99, 67

\bibitem[\protect\citeauthoryear{{Neufeld}, {Maloney}, \& {Conger}}{{Neufeld}
  et~al.}{1994}]{NMC94}
{Neufeld}, D.~A., {Maloney}, P.~R.,  \& {Conger}, S. 1994, ApJ, 436, L127

\bibitem[\protect\citeauthoryear{{Norman}, {Sellwood}, \& {Hasan}}{{Norman}
  et~al.}{1996}]{NSH96}
{Norman}, C.~A., {Sellwood}, J.~A.,  \& {Hasan}, H. 1996, ApJ, 462, 114

\bibitem[\protect\citeauthoryear{{O'Dea}, {Baum}, \& {Gallimore}}{{O'Dea}
  et~al.}{1994}]{OBG94}
{O'Dea}, C.~P., {Baum}, S.~A.,  \& {Gallimore}, J.~F. 1994, ApJ, 436, 669

\bibitem[\protect\citeauthoryear{{O'Dea}, {Baum}, \& {Stanghellini}}{{O'Dea}
  et~al.}{1991}]{OBS91}
{O'Dea}, C.~P., {Baum}, S.~A.,  \& {Stanghellini}, C. 1991, ApJ, 380, 66

\bibitem[\protect\citeauthoryear{{Payne}, {Salpeter}, \& {Terzian}}{{Payne}
  et~al.}{1982}]{PST82}
{Payne}, H.~E., {Salpeter}, E.~E.,  \& {Terzian}, Y. 1982, ApJS, 48, 199

\bibitem[\protect\citeauthoryear{{Payne}, {Salpeter}, \& {Terzian}}{{Payne}
  et~al.}{1983}]{PST83}
{Payne}, H.~E., {Salpeter}, E.~E.,  \& {Terzian}, Y. 1983, ApJ, 272, 540

\bibitem[\protect\citeauthoryear{{Pedlar} et~al.}{{Pedlar}
  et~al.}{1998}]{PFHRD98}
{Pedlar}, A., {Fernandez}, B., {Hamilton}, N.~G., {Redman}, M.~P.,  \&
  {Dewdney}, P.~E. 1998, MNRAS, 300, 1071

\bibitem[\protect\citeauthoryear{{Pedlar} et~al.}{{Pedlar}
  et~al.}{1992}]{PHAU92}
{Pedlar}, A., {Howley}, P., {Axon}, D.~J.,  \& {Unger}, S.~W. 1992, MNRAS, 259,
  369

\bibitem[\protect\citeauthoryear{{Pedlar} et~al.}{{Pedlar}
  et~al.}{1993}]{PKLMABOU93}
{Pedlar}, A., {Kukula}, M., {Longley}, D. P.~T., {Muxlow}, T. W.~B., {Axon},
  D.~J., {Baum}, S., {O'Dea}, C.~P.,  \& {Unger}, S.~W. 1993, MNRAS, 263, 471

\bibitem[\protect\citeauthoryear{{Pedlar} et~al.}{{Pedlar}
  et~al.}{1996}]{PMGBO96}
{Pedlar}, A., {Mundell}, C.~G., {Gallimore}, J.~F., {Baum}, S.~A.,  \& {O'Dea},
  C.~P. 1996, Vistas in Astronomy, 40, 91

\bibitem[\protect\citeauthoryear{{Peterson}}{{Peterson}}{1997}]{Peterson97}
{Peterson}, B.~M. 1997, An Introduction to Active Galactic Nuclei (Cambridge:
  Cambridge University Press)

\bibitem[\protect\citeauthoryear{{Phinney}}{{Phinney}}{1994}]{Phinney94p1}
{Phinney}, E.~S. 1994, in Mass Transfer Induced Activity in Galaxies, ed.
  I.~{Shlosman} (Cambridge: Cambridge University Press), 1

\bibitem[\protect\citeauthoryear{{Press} et~al.}{{Press} et~al.}{1992}]{PTVF92}
{Press}, W.~H., {Teukolsky}, S.~A., {Vetterling}, W.~T.,  \& {Flannery}, B.~P.
  1992, Numerical Recipes, second edition (Cambridge: Cambridge University
  Press)

\bibitem[\protect\citeauthoryear{{Read}, {Ponman}, \& {Strickland}}{{Read}
  et~al.}{1997}]{RPS97}
{Read}, A.~M., {Ponman}, T.~J.,  \& {Strickland}, D.~K. 1997, MNRAS, 286, 626

\bibitem[\protect\citeauthoryear{{Rees}}{{Rees}}{1984}]{Rees84}
{Rees}, M.~J. 1984, ARAA, 22, 471

\bibitem[\protect\citeauthoryear{{Rigopoulou} et~al.}{{Rigopoulou}
  et~al.}{1996}]{RLWRC96}
{Rigopoulou}, D., {Lawrence}, A., {White}, G.~J., {Rowan-Robinson}, M.,  \&
  {Church}, S.~E. 1996, A\&A, 305, 747

\bibitem[\protect\citeauthoryear{{Roberts}}{{Roberts}}{1975}]{Roberts75p309}
{Roberts}, M.~S. 1975, in Galaxies and the Universe, ed. A.~{Sandage},
  M.~{Sandage}, \& J.~{Kristian} (Chicago: University of Chicago Press), 309

\bibitem[\protect\citeauthoryear{{Roberts}, {Huntley}, \& {van
  Albada}}{{Roberts} et~al.}{1979}]{RHvA79}
{Roberts}, W. W.~J., {Huntley}, J.~M.,  \& {van Albada}, G.~D. 1979, ApJ, 233,
  67

\bibitem[\protect\citeauthoryear{{Roy} et~al.}{{Roy} et~al.}{1998}]{RCWU98}
{Roy}, A.~L., {Colbert}, E. J.~M., {Wilson}, A.~S.,  \& {Ulvestad}, J.~S. 1998,
  ApJ, 504, 147

\bibitem[\protect\citeauthoryear{{Rubin}}{{Rubin}}{1978}]{Rubin78}
{Rubin}, V.~C. 1978, ApJ, 224, L55

\bibitem[\protect\citeauthoryear{{Saikia} et~al.}{{Saikia}
  et~al.}{1994}]{SPUA94}
{Saikia}, D.~J., {Pedlar}, A., {Unger}, S.~W.,  \& {Axon}, D.~J. 1994, MNRAS,
  270, 46

\bibitem[\protect\citeauthoryear{{Sandage} \& {Tammann}}{{Sandage} \&
  {Tammann}}{1981}]{ST81}
{Sandage}, A.,  \& {Tammann}, G.~A. 1981, Revised Shapley-Ames Catalog of
  Bright Galaxies (Washington: Carnegie Institution of Washington)

\bibitem[\protect\citeauthoryear{{Sanders} et~al.}{{Sanders}
  et~al.}{1989}]{SPNSM89}
{Sanders}, D.~B., {Phinney}, E.~S., {Neugebauer}, G., {Soifer}, B.~T.,  \&
  {Matthews}, K. 1989, ApJ, 347, 29

\bibitem[\protect\citeauthoryear{{Sanders} \& {Tubbs}}{{Sanders} \&
  {Tubbs}}{1980}]{ST80}
{Sanders}, R.~H.,  \& {Tubbs}, A.~D. 1980, ApJ, 235, 803

\bibitem[\protect\citeauthoryear{{Satoh}, {Inoue}, \& {Nakai}}{{Satoh}
  et~al.}{1998}]{SIN98p219}
{Satoh}, S., {Inoue}, M.,  \& {Nakai}, N. 1998, in Radio Emission from Galactic
  and Extragalactic Compact Sources, ed. J.~A. {Zensus}, G.~B. {Taylor}, \&
  J.~M. {Wrobel} (San Francisco: ASP), 219

\bibitem[\protect\citeauthoryear{{Schmitt} et~al.}{{Schmitt}
  et~al.}{1997}]{SKSA97}
{Schmitt}, H.~R., {Kinney}, A.~L., {Storchi-Bergmann}, T.,  \& {Antonucci}, R.
  1997, ApJ, 422, 623

\bibitem[\protect\citeauthoryear{{Schwarz}, {Ekers}, \& {Goss}}{{Schwarz}
  et~al.}{1982}]{SEG82}
{Schwarz}, U.~J., {Ekers}, R.~D.,  \& {Goss}, W.~M. 1982, A\&A, 110, 100

\bibitem[\protect\citeauthoryear{{S\'ersic}}{{S\'ersic}}{1973}]{Sersic73}
{S\'ersic}, J.~L. 1973, PASP, 85, 103

\bibitem[\protect\citeauthoryear{{Shlosman}, {Frank}, \& {Begelman}}{{Shlosman}
  et~al.}{1989}]{SFB89}
{Shlosman}, I., {Frank}, J.,  \& {Begelman}, M.~C. 1989, Nature, 338, 45

\bibitem[\protect\citeauthoryear{{Shostak} \& {van der Kruit}}{{Shostak} \&
  {van der Kruit}}{1984}]{SvdK84}
{Shostak}, S.,  \& {van der Kruit}, P.~C. 1984, A\&A, 132, 20

\bibitem[\protect\citeauthoryear{{Simcoe} et~al.}{{Simcoe}
  et~al.}{1997}]{SMSE97}
{Simcoe}, R., {McLeod}, K.~K., {Schachter}, J.,  \& {Elvis}, M. 1997, ApJ, 489,
  615

\bibitem[\protect\citeauthoryear{{Simkin} et~al.}{{Simkin}
  et~al.}{1987}]{SvGHS87}
{Simkin}, S.~M., {van Gorkom}, J., {Hibbard}, J.,  \& {Su}, H.-J. 1987,
  Science, 235, 1367

\bibitem[\protect\citeauthoryear{{Smith} \& {Done}}{{Smith} \&
  {Done}}{1996}]{SD96}
{Smith}, D.,  \& {Done}, C. 1996, MNRAS, 280, 355

\bibitem[\protect\citeauthoryear{{Stanghellini} et~al.}{{Stanghellini}
  et~al.}{1997}]{SBDOBFF97}
{Stanghellini}, C., {Bondi}, M., {Dallacasa}, D., {O'Dea}, C.~P., {Baum},
  S.~A., {Fanti}, R.,  \& {Fanti}, C. 1997, A\&A, 318, 376

\bibitem[\protect\citeauthoryear{{Thuan} \& {Wadiak}}{{Thuan} \&
  {Wadiak}}{1982}]{TW82}
{Thuan}, T.~X.,  \& {Wadiak}, E.~J. 1982, ApJ, 252, 125

\bibitem[\protect\citeauthoryear{{Toomre}}{{Toomre}}{1964}]{Toomre64}
{Toomre}, A. 1964, ApJ, 139, 1217

\bibitem[\protect\citeauthoryear{{Tran}, {Miller}, \& {Kay}}{{Tran}
  et~al.}{1992}]{TMK92}
{Tran}, H.~D., {Miller}, J.~S.,  \& {Kay}, L.~E. 1992, ApJ, 397, 457

\bibitem[\protect\citeauthoryear{{Turner} et~al.}{{Turner}
  et~al.}{1997}]{TGNM97}
{Turner}, T.~J., {George}, I.~M., {Nandra}, K.,  \& {Mushotzky}, R.~F. 1997,
  ApJS, 113, 23

\bibitem[\protect\citeauthoryear{{Ulvestad} et~al.}{{Ulvestad}
  et~al.}{1998}]{URCW98}
{Ulvestad}, J.~S., {Roy}, A.~L., {Colbert}, E. J.~M.,  \& {Wilson}, A.~S. 1998,
  ApJ, 496, 196

\bibitem[\protect\citeauthoryear{{Ulvestad} \& {Wilson}}{{Ulvestad} \&
  {Wilson}}{1984}]{UW84b}
{Ulvestad}, J.~S.,  \& {Wilson}, A.~S. 1984, ApJ, 285, 439

\bibitem[\protect\citeauthoryear{{Ulvestad} \& {Wilson}}{{Ulvestad} \&
  {Wilson}}{1989}]{UW89}
{Ulvestad}, J.~S.,  \& {Wilson}, A.~S. 1989, ApJ, 343, 659

\bibitem[\protect\citeauthoryear{{Ulvestad} et~al.}{{Ulvestad}
  et~al.}{1999}]{UWRWFK99}
{Ulvestad}, J.~S., {Wrobel}, J.~M., {Roy}, A.~L., {Wilson}, A.~S., {Falcke},
  H.,  \& {Kritchbaum}, T.~P. 1999, ApJ, in press

\bibitem[\protect\citeauthoryear{{Unger} et~al.}{{Unger} et~al.}{1986}]{UPBH86}
{Unger}, S.~W., {Pedlar}, A., {Booler}, R.~V.,  \& {Harrison}, B.~A. 1986,
  MNRAS, 219, 387

\bibitem[\protect\citeauthoryear{{Unger} et~al.}{{Unger}
  et~al.}{1984}]{UPNdB84}
{Unger}, S.~W., {Pedlar}, A., {Neff}, S.~G.,  \& {de Bruyn}, A.~G. 1984, MNRAS,
  209, 15p

\bibitem[\protect\citeauthoryear{{van der Hulst}, {Hummel}, \& {Dickey}}{{van
  der Hulst} et~al.}{1982}]{vdHHD82}
{van der Hulst}, J.~M., {Hummel}, E.,  \& {Dickey}, J.~M. 1982, ApJ, 261, L59

\bibitem[\protect\citeauthoryear{{van Driel}}{{van Driel}}{1987}]{drielthesis}
{van Driel}, W. 1987, Ph. D. Thesis (Groningen, the Netherlands: University of
  Groningen)

\bibitem[\protect\citeauthoryear{{van Gorkom,} et~al.}{{van Gorkom,}
  et~al.}{1989}]{vGKEELP89}
{van Gorkom,}, J.~H., {Knapp}, G.~R., {Ekers}, R.~D., {Ekers}, D.~D., {Laing},
  R.~A.,  \& {Polk}, K.~S. 1989, AJ, 97, 708

\bibitem[\protect\citeauthoryear{{van Woerden}, {van Driel}, \&
  {Schwartz}}{{van Woerden} et~al.}{1983}]{vW+83}
{van Woerden}, H., {van Driel}, W.,  \& {Schwartz}, U.~J. 1983, in Internal
  Kinematics and Dynamics of Galaxies, ed. E.~{Athanassoula} (Dordrecht:
  Reidel), 99

\bibitem[\protect\citeauthoryear{{Wada} \& {Habe}}{{Wada} \&
  {Habe}}{1992}]{WH92}
{Wada}, K.,  \& {Habe}, A. 1992, MNRAS, 258, 82

\bibitem[\protect\citeauthoryear{{Wang}, {Brinkmann}, \& {Bergeron}}{{Wang}
  et~al.}{1996}]{WBB96}
{Wang}, T., {Brinkmann}, W.,  \& {Bergeron}, J. 1996, A\&A, 309, 81

\bibitem[\protect\citeauthoryear{{Ward} et~al.}{{Ward} et~al.}{1980}]{WPBT80}
{Ward}, M., {Penston}, M.~V., {Blades}, J.~C.,  \& {Turtle}, A.~J. 1980, MNRAS,
  193, 563

\bibitem[\protect\citeauthoryear{{Ward}}{{Ward}}{1996}]{Ward96}
{Ward}, M.~J. 1996, Vistas in Astronomy, 40, 233

\bibitem[\protect\citeauthoryear{{Wehrle} \& {Morris}}{{Wehrle} \&
  {Morris}}{1987}]{WM87}
{Wehrle}, A.~E.,  \& {Morris}, M. 1987, ApJ, 313, L43

\bibitem[\protect\citeauthoryear{{Wehrle} \& {Morris}}{{Wehrle} \&
  {Morris}}{1988}]{WM88}
{Wehrle}, A.~E.,  \& {Morris}, M. 1988, AJ, 95, 1689

\bibitem[\protect\citeauthoryear{{Whittle}}{{Whittle}}{1992}]{Whittle92a}
{Whittle}, M. 1992, ApJS, 79, 49

\bibitem[\protect\citeauthoryear{{Wilson}}{{Wilson}}{1991}]{Wilson91p227}
{Wilson}, A. 1991, in The Interpretation of Modern Synthesis Observations of
  Spiral Galaxies, ed. N.~{Duric} \& P.~C. {Crane} (San Francisco: ASP), 227

\bibitem[\protect\citeauthoryear{{Wilson} \& {Baldwin}}{{Wilson} \&
  {Baldwin}}{1985}]{WB85}
{Wilson}, A.~S.,  \& {Baldwin}, J.~A. 1985, ApJ, 289, 124

\bibitem[\protect\citeauthoryear{{Wilson}, {Braatz}, \& {Henkel}}{{Wilson}
  et~al.}{1995}]{WBH95}
{Wilson}, A.~S., {Braatz}, J.~A.,  \& {Henkel}, C. 1995, ApJ, 455, L127

\bibitem[\protect\citeauthoryear{{Wilson} et~al.}{{Wilson}
  et~al.}{1976}]{WPFB76}
{Wilson}, A.~S., {Penston}, M.~V., {Fosbury}, R. A.~E.,  \& {Boksenberg}, A.
  1976, MNRAS, 177, 673

\bibitem[\protect\citeauthoryear{{Wilson} \& {Ulvestad}}{{Wilson} \&
  {Ulvestad}}{1983}]{WU83}
{Wilson}, A.~S.,  \& {Ulvestad}, J.~S. 1983, ApJ, 275, 8

\end{thebibliography}
\end{document}